\documentclass[journal,draftclsnofoot,onecolumn,12pt,twoside]{IEEEtran}
\usepackage{graphicx}
\usepackage{amsmath,amssymb}
\usepackage{cases}
\usepackage{color}
\usepackage{amsfonts}
\usepackage{indentfirst}
\usepackage{cite}
\usepackage[justification=centering]{caption}
\usepackage{diagbox}
\usepackage{bm}
\usepackage{amssymb}
\usepackage{caption}
\usepackage{algorithm}
\usepackage{algpseudocode}
\usepackage{booktabs}
\usepackage{subfigure}
\usepackage{multirow}
\usepackage{amsthm}
\usepackage{chngpage}
\usepackage{epstopdf}
\usepackage[hidelinks]{hyperref}
\usepackage{pstricks}
\usepackage{blkarray}
\usepackage{arydshln}
\usepackage{amsmath}
\makeatletter
\newtheoremstyle{mythm}{3pt}{3pt}{}{16pt}{\bfseries}{:}{.5em}{}
\theoremstyle{mythm}
\newtheorem{theorem}{Theorem}
\newtheorem{example}{Example}
\newtheorem{definition}{Definition}

\newtheorem{corollary}{Corollary}
\newtheorem{lemma}{Lemma}
\newtheorem{construction}{Construction}

\newcommand{\tabincell}[2]{\begin{tabular}{@{}#1@{}}#2\end{tabular}}
\begin{document}
\title{Coded Caching with Linear Subpacketization via Hamming Distance and Linear Codes}
\author{Jinyu~Wang, Guangyu~Pang, Minquan~Cheng,~\IEEEmembership{Member,~IEEE,} Jiahui~Wu, Xingjian~Cao
	\thanks{Jinyu~Wang, Guangyu~Pang, and Jiahui~Wu are with School of Mathematics and Statistics, Guangxi Normal University, 541004 Guilin, China (e-mail: mathwjy@163.com, 1138629317@qq.com, 2026344167@qq.com).}
	
	\thanks{Minquan Cheng is with the Key Lab of Education Blockchain and Intelligent Technology, Ministry of Education, and also with the Guangxi Key Lab of Multi-source Information Mining $\&$ Security, Guangxi Normal University, 541004 Guilin, China (e-mail: chengqinshi@hotmail.com).}
	
	\thanks{Xingjian~Cao is with School of Mathematics and Physics, Xi'an Jiaotong-Liverpool University, 215123 Suzhou, China (e-mail: Xingjian.Cao25@student.xjtlu.edu.cn). }
}

\date{}
\maketitle

\begin{abstract}
	Coded caching exploits users' local caches to reduce communication load, while placement delivery arrays (PDAs) provide a unified combinatorial framework for representing uncoded placement and one-shot linear delivery. PDA-based coded caching schemes with linear subpacketization, particularly those with subpacketization $F$ equal to the number of users $K$, have attracted considerable attention due to their potential for reducing implementation complexity. However, many existing constructions impose a regularity constraint on PDAs, requiring every non-star symbol to occur the same number of times. Under the linear-subpacketization constraint, this regularity requirement may limit the achievable transmission performance.
	
	In this paper, we propose two classes of PDA constructions based on Hamming distance and linear codes, both achieving \(F=K\). The first construction places non-star entries only at positions with a prescribed Hamming distance and employs auxiliary linear codes to obtain large symbol multiplicities. When coordinate alphabet sizes are not all equal, the resulting symbol multiplicities may vary, leading to different coded multicast gains. When all coordinate alphabets have the same size, we consider two alternative symbol assignments based on binary linear codes and \(\mathbb{Z}_q\)-linear codes, respectively, and select the one with the larger coded multicast gain. The second construction places non-star entries in consecutive Hamming-distance layers and independently employs suitable binary linear codes in each layer, allowing different layers to achieve different coded multicast gains. 
	The theoretical and numerical comparisons show that the proposed schemes can achieve lower transmission loads than several existing linear-subpacketization schemes at the same or comparable memory ratios. Moreover, they achieve substantial reductions in subpacketization compared with schemes having other subpacketization levels, while maintaining competitive transmission loads.
\end{abstract}
\begin{IEEEkeywords}
	Coded caching, placement delivery array, linear subpacketization, Hamming distance, linear code.
\end{IEEEkeywords}

\section{Introduction}\label{sec:introduction}

The growing demand for video streaming and other data-intensive services puts increasing pressure on communication networks. Network traffic also varies over time, leading to congestion during peak times and underutilization of resources during off-peak times. Caching exploits this temporal variation by placing content in users' caches during off-peak times, thereby reducing the traffic that must be delivered during peak times. Conventional caching provides a local caching gain, as users can obtain part of the requested content directly from their own caches. Maddah-Ali and Niesen (MN) showed that coordinating cache placement and coded delivery can provide an additional global caching gain~\cite{MN}. By combining packets intended for different users in a single transmission, coded caching allows the server to serve multiple users simultaneously.

Consider a centralized coded caching system in which a server stores $N$ equal-sized files and serves $K$ users through a shared broadcast link. Each user has a cache of size $M$ files. The operation consists of a placement phase and a delivery phase. During placement, each file is divided into $F$ equal-sized packets, and the caches are populated before the users' demands are known. During delivery, the server broadcasts coded messages so that each user can recover its requested file using its cache contents. Here, $F$ is the subpacketization, and the transmission load $R$ is defined as the maximum amount of transmitted data over all demands, normalized by the size of one file.

For \(K\leq N\), the MN scheme achieves a transmission load within a factor of \(2\) of the information-theoretic lower bound \cite{YMA2019} and is optimal among schemes with uncoded placement~\cite{WTP2016,YMA2018,WTP2020}. However, its subpacketization grows exponentially with the number of users, increasing the complexity of placement and delivery and hindering practical implementation. To address this issue, Shanmugam et al. \cite{SJTLD} explored user grouping, while Yan et al. \cite{YCTC} introduced the placement delivery array (PDA) framework. A PDA is an \(F\times K\) array consisting of stars and non-star entries, where a star in row \(f\) and column \(k\) indicates that the \(k\)-th user caches the \(f\)-th packet of every file, while a non-star entry indicates that the corresponding packet is not cached by that user. During delivery, the server XORs the requested packets associated with the same non-star symbol and broadcasts the resulting coded message. The number of occurrences of a non-star symbol determines its coded multicast gain, i.e., the number of users served simultaneously. This combinatorial framework provides a systematic way to characterize the tradeoff between transmission load and subpacketization. Since then, considerable efforts have been devoted to developing PDA-based schemes with reduced subpacketization.

Specifically, Yan et al.~\cite{YCTC} achieve asymptotically optimal load, although only for certain memory ratios. Shanmugam et al.~\cite{SZG} employ $(6,3)$-free hypergraphs to support flexible numbers of users with sub-exponential subpacketization. The work in~\cite{TR} generalizes the construction in~\cite{YCTC} using linear block codes, while~\cite{CWZW} provides a unified framework encompassing the MN scheme and the constructions in~\cite{YCTC,SZG} through flexible selections of row and column sets. Furthermore,~\cite{CJYT} extends the constructions in~\cite{YCTC,SZG} to more general memory ratios, whereas~\cite{WCLC} employs orthogonal arrays to generalize or improve several existing constructions in~\cite{YCTC,CJYT,CWZW,SZG}.
In addition, various combinatorial structures and construction techniques have been exploited in coded caching, including, but not limited to, strong edge colorings of bipartite graphs \cite{YTCC,MW}, $t$-designs \cite{ASK,LC_sharelink,TRR}, projective geometry \cite{CKSM}, Cartesian-product constructions \cite{WCWC2023,WCWC2026}, rainbow arithmetic progression sets \cite{XXGL},  Hamming-distance-based constructions \cite{ZCW}, injective arc colorings of regular digraphs \cite{WCCLS}, lifting techniques \cite{AST}, and non-half-sum combinatorial structures \cite{CWWC,YCWC}. Table~\ref{tab_general} summarizes the parameters of these existing schemes.

\begin{table*}
	\centering
	\caption{Parameters of Existing Coded Caching Schemes. Define $\left[k \atop t\right]_q=\frac{(q^k-1)\cdots(q^{k-t+1}-1)}{(q^t-1)\cdots(q-1)}$.}
	\label{tab_general}
	\begin{tabular}{cccccc}
		\hline
		Scheme & Users $K$ & Memory ratio $M/N$ & Load $R$ & Subpacketization $F$ & Constraints \\
		\hline
		MN~\cite{MN} & $K$ & $\frac{t}{K}$ & $\frac{K - t}{t + 1}$ & $\binom{K}{t}$ & $K\in \mathbb{Z}^+$, $t\in[0:K]$ \\
		\hline
		
		YTCC~\cite{YTCC} & $\binom{H}{a}$ & $1 - \frac{\binom{a}{r}\binom{H - a}{b - r}}{\binom{H}{b}}$ & \tabincell{c}{$\frac{{H\choose a+b-2r}}{{H\choose b}}\cdot$\\ $\min\{{H-a-b+2r\choose r},$\\ ${{a+b-2r\choose a-r}}\}$} & $\binom{H}{b}$ & \tabincell{c}{$H,a,b,r\in \mathbb{Z}$, \\
			$0\leq r<a<H$, \\$r<b<H$\\ $a+b\leq H+r$}
		\\
		\hline
		
		WCLC~\cite{WCLC} & $\binom{m}{t}q^{t}$ & $1 - \left( \frac{q - z}{q} \right)^{t}$ & $\left( \frac{q - z}{\lfloor\frac{q-1}{q-z}\rfloor} \right)^{t}$ & $\lfloor\frac{q-1}{q-z}\rfloor^{t}q^{m - 1}$ & \tabincell{c}{$m,t,q,z\in \mathbb{Z}^+$,\\ $z < q$, $t\leq m$} \\
		\hline
		
		
		CKSM 1~\cite{CKSM} &  \tabincell{c}{$\frac{1}{t!}q^{\frac{t(t-1)}{2}}$\\$\prod_{i=0}^{t-1}\left[k-i \atop 1\right]_q$}& \tabincell{c}{$1-q^{mt}$\\$\prod_{i=0}^{m-1}\frac{\left[k-t-i \atop 1\right]_q}{\left[k-i \atop 1\right]_q}$}  & \tabincell{c}{$\frac{m!q^{mt}}{(m+t)!}q^{\frac{t(t-1)}{2}}$\\$\prod_{i=0}^{t-1}\left[k-m-i \atop 1\right]_q$}& \tabincell{c}{$\frac{1}{m!}q^{\frac{m(m-1)}{2}}$\\$\prod_{i=0}^{m-1}\left[k-i \atop 1\right]_q$}&\multirow{2}{*}{\tabincell{c}{$k,m,t\in\mathbb{Z}^{+}$,\\ $m+t\leq k$;\\ $q$ is a prime power}}\\ 
		\cline{1-5}	
			
		CKSM 2~\cite{CKSM} & $\genfrac{[}{]}{0pt}{}{k}{t}_{q}$ & $1 - \frac{\genfrac{[}{]}{0pt}{}{k - t}{m}_{q}}{\genfrac{[}{]}{0pt}{}{k}{m + t}_{q}}$ & $\frac{\genfrac{[}{]}{0pt}{}{k}{m}_{q}}{\genfrac{[}{]}{0pt}{}{k}{m + t}_{q}}$ & $\genfrac{[}{]}{0pt}{}{k}{m + t}_{q}$ &  \\
		\hline

		ZCW 1~\cite{ZCW} & $2^m$ & $1-\frac{\binom{m}{w}}{\sum_{i=0}^{w}\binom{m}{i}}$ & $\frac{\binom{m}{w}2^{m-w}}{\sum_{i=0}^{w}\binom{m}{i}}$
		& $\sum_{i=0}^{w}\binom{m}{i}$ &\multirow{2}{*}{\tabincell{c}{$m,w\in \mathbb{Z}^+$,\\ $w<m$}} \\
		\cline{1-5}	
			
		ZCW 2~\cite{ZCW} & $3^{m}$ & $1-\frac{\binom{m}{w}2^w}{\sum_{i=0}^{w}\binom{m}{i}2^i}$ & $\frac{\binom{m}{w}3^m}{\sum_{i=0}^{w}\binom{m}{i}2^i}$ & $\sum_{i=0}^{w}\binom{m}{i}2^i$ & \\
		\hline
		
		WCCLS~\cite{WCCLS} & $q^{m}$ & $1 - \frac{\binom{m}{w}(q - 1)^{w}}{q^{m}}$ & $\frac{\binom{m}{w}(q - 1)^{w}}{q^{m - w}}$ & $q^{m}$ & \tabincell{c}{$m,w\in \mathbb{Z}^+$, \\$w < m$, $q \geq 2$}\\
		\hline		
		
		ASK 1~\cite{ASK} & $q^{2} + q + 1$ & $\frac{q^{2}}{q^{2} + q + 1}$ & $1$ & $q^{2} + q + 1$ & \multirow{2}{*}{$q$ is a prime power} \\
		\cline{1-5}
		ASK 2~\cite{ASK} & $q^{2}$ & $\frac{q - 1}{q}$ & $\frac{q}{q + 1}$ & $q^{2} + q$ &   \\
		\hline
		
		AST 1~\cite{AST} & $g^2$ & $1-\frac{2}{g}+\frac{1}{g^2}$ & $2-\frac{1}{g}$ & $g^2$ & \multirow{2}{*}{$g\in\mathbb{Z}^+$, $g \geq 2$} \\
		\cline{1-5}
		
		AST 2~\cite{AST} & $2g^2$ & $1-\frac{3}{2g}+\frac{1}{g^2}$ & $\frac{3}{2}-\frac{1}{g}$ & $2g^2$ & \\
		\hline
		
		AST 3~\cite{AST} & $2gd$ & $\frac{2gd-3g+\frac{g}{d}+1}{2gd}$ & $\frac{3g-\frac{g}{d}-1}{2d}$ & $2gd$ & $d,g\in \mathbb{Z}^+$, $d|g$ \\
		\hline
		
		AST 4~\cite{AST} & $2gd$ & $\frac{2gd-3g+\frac{g}{d}+2d-d^2}{2gd}$ & $\frac{3g-\frac{g}{d}-2d+d^2}{2d}$ & $2gd$ & $d,g\in \mathbb{Z}^+$, $d^2|g$ \\
		\hline
		
		AST 5~\cite{AST} & $gd$ & $\frac{gd+d-2g+\frac{g}{d}}{gd}$ & $\frac{2g-d-\frac{g}{d}}{2d}$ & $gd$ & $d,g\in \mathbb{Z}^+$, $d|g$ \\
		\hline
		
		AST 6~\cite{AST} & $4n^3$ & $\frac{4n^3-8n^2+7n-2}{4n^3}$ & $\frac{8n^2-7n+2}{2n}$ & $4n^3$ & $n\in \mathbb{Z}^+$ \\
		\hline
		
		AST 7~\cite{AST} & $bg$ & $1-\frac{b+d-1}{bg}$ & $\frac{b+d-1}{g}$ & $bg$ & \tabincell{c}{$g,b\in \mathbb{Z}^+$,\\ $d=\gcd(g,b)$} \\
		\hline
		
		AST 8~\cite{AST} & $2^{r}k$ & $1 - \frac{(r + 1)k - r}{2^{r}k}$ & $\frac{(r + 1)k - r}{2^{r}}$ & $2^{r}k$ & $r,k \in \mathbb{Z}^{+}$ \\
		\hline
		
		CWWC~\cite{CWWC} & $v$ & $1 - \frac{2^{n}\lfloor\frac{v^{1/n} - 1}{2}\rfloor^{n}}{v}$ & $\lfloor\frac{v^{1/n} - 1}{2}\rfloor^{n}$ & $v$ &\tabincell{c}{ $v,n\in \mathbb{Z}^{+}$, $v$ is odd;\\ $v^{1/n}\geq 3$ }\\
		\hline
		
		YCWC~\cite{YCWC} & $v$ & $1 -\frac{\lfloor v^{1/n} - 1\rfloor^n}{v}$ & $\frac{\lfloor v^{1/n} - 1\rfloor^n}{2^n}$ & $2^nv$ &\tabincell{c}{ $v,n\in \mathbb{Z}^{+}$, $v$ is odd;\\ $v^{1/n}\geq 2$} \\
		\hline
	\end{tabular}
	
\end{table*}

In recent years, coded caching schemes with linear subpacketization, particularly those with \(F=K\), have attracted increasing attention. Existing schemes can be broadly classified into two categories. The first category determines the cache placement based on specific mathematical structures and then designs the delivery strategy with the aim of maximizing the number of users served by each multicast transmission, thereby pursuing a large coded multicast gain; see, e.g., \cite{ASK,AST,ZCW,WCCLS}. The second category first specifies a delivery strategy and then designs a corresponding cache placement with the aim of minimizing the cache memory required at each user; see, e.g., \cite{CWWC,YCWC}. Most of these schemes share a common feature: every multicast transmission serves the same number of users, which corresponds to the regularity property that every non-star symbol in the associated PDA occurs the same number of times.
However, the design of delivery strategies is already highly constrained under linear subpacketization, particularly when \(F=K\). Requiring every multicast transmission to serve the same number of users imposes an additional restriction, further limiting the achievable transmission performance. Motivated by this observation, we propose two classes of PDA constructions based on Hamming distance and linear codes. The key idea is to use Hamming-distance constraints to construct the cache placement and classify each user's missing packets, and then employ linear codes to encode packets within each class for coded multicasting.
Unlike existing regular PDA-based schemes, our constructions allow different multicast transmissions to have different multicast gains, thereby providing greater flexibility in improving transmission performance.

\subsection{Contributions}
We propose two classes of PDA constructions based on Hamming distance. Specifically, we use \(m\)-dimensional vectors
$\mathbf{a}=(a_1,\ldots,a_m)$ (where $a_i\in \{0,1,\ldots,q_i-1\}$ and each $q_i\geq 2$ is an integer) 
as both the row and column labels of the PDA, thereby achieving \(F=K=\prod_{i=1}^{m}q_i\). The array positions are partitioned into layers according to the Hamming distance between their row and column labels. The main contributions are summarized as follows.

\noindent
{\bf 1) Single-Hamming-distance-based PDA construction:} For any \(w\in\{0,1,\ldots,m\}\), we place non-star entries at positions where the Hamming distance between the row and column labels is exactly \(w\), and stars elsewhere. The non-star symbols are assigned using a binary linear code $\mathcal{C}$ of length \(w\), minimum distance at least \(m-w+1\), and maximum possible dimension \(\delta\), aiming to maximize the multiplicity of each non-star symbol.

When \(q_1=\cdots=q_m=q\), all non-star positions belong to the same class, and each non-star symbol occurs exactly \(2^\delta q^{m-w}\) times in the above construction. Alternatively, for the same placement, we can employ a linear code over $\mathbb{Z}_q$ of length $m-w$ with the maximum possible cardinality $\nu=|\mathcal{C}'|$ subject to a minimum Hamming distance of at least \(w+1\) . This yields \(2^w\nu\) occurrences per symbol. By choosing the better of the two assignments, we obtain a coded caching scheme with \(F=K=q^m\) and coded multicast gain
$
g=\max\left\{2^\delta q^{m-w},\,2^w\nu\right\}$.

\noindent
\textbf{2) Consecutive-Hamming-distance-based PDA construction:}
Since the first construction places non-star entries only in a single Hamming-distance layer, it may result in a relatively high memory ratio. To achieve linear-subpacketization schemes over a broader range of memory ratios, we propose a second construction. For any $w\in\{1,\ldots,m\}$, an entry is a star if the Hamming distance between its row and column labels is less than $w$, and a non-star otherwise. For each $r\in\{w,\ldots,m\}$, we use a binary linear code of length $r$, minimum distance at least $r-w+1$, and maximum possible dimension $\delta_r$ to assign symbols to non-star entries in the corresponding Hamming-distance layer, aiming to maximize the occurrence of each symbol within that layer. Since $\delta_r$ may vary with $r$, the symbol multiplicities may also differ across layers, even when all coordinate alphabets have the same size. Thus, rather than imposing a uniform coded-multicast gain across all transmissions, we independently select a suitable linear code for each layer according to its local structure, allowing potentially large but not necessarily equal coded-multicast gains and thereby reducing the overall transmission load.

\noindent
{\bf 3) Performance analysis:} We conduct theoretical and numerical comparisons with existing coded caching schemes having different subpacketization levels. The theoretical analysis identifies conditions under which the proposed schemes achieve lower transmission loads than WCCLS~\cite{WCCLS} and CWWC~\cite{CWWC}, while maintaining \(F=K\). Numerical results show that, for the same number of users and at the same or comparable memory ratios, the proposed schemes can achieve lower transmission loads than existing linear-subpacketization schemes, including AST~\cite{AST}, ZCW~\cite{ZCW}, WCCLS~\cite{WCCLS}, ASK~\cite{ASK}, CWWC~\cite{CWWC}, and YCWC~\cite{YCWC}. Comparisons with schemes having other subpacketization levels, including MN~\cite{MN}, CKSM~\cite{CKSM}, WCLC~\cite{WCLC}, and YTCC~\cite{YTCC}, further highlight the advantages of the proposed schemes. Specifically, compared with MN, the proposed schemes reduce the subpacketization from exponential to linear, i.e., from \(\binom{K}{t}\) to \(K\). They also achieve both lower transmission loads and smaller subpacketization than CKSM over a wide range of memory ratios. Compared with WCLC, they achieve lower transmission loads over a broad range of memory ratios while maintaining comparable subpacketization.

The rest of this paper is organized as follows. Section~\ref{sec:preliminaries} introduces the system model and preliminaries. Section~\ref{sec:constructions} presents the proposed PDA constructions and main results. Section~\ref{sec:performance} analyzes the performance of the proposed schemes. Finally, Section~\ref{sec:conclusion} concludes the paper, with some proofs deferred to the appendices.

\noindent
\textbf{Notation:} For integers $m\leq n$, let $[m:n]=\{m,m+1,\ldots,n\}$ and $[m]=\{1,2,\ldots,m\}$. The cardinality of a set $\mathcal{A}$ is denoted by $|\mathcal{A}|$.

\section{Preliminaries}\label{sec:preliminaries}
In this section,we introduce the system model considered in this paper and some related works.

\subsection{System Model}

Consider a centralized coded caching system in which a server stores $N$ equal-sized files, denoted by $\mathcal{W}=\{W_n|n\in[N]\}$, and serves $K$ users through an error-free shared link satisfying $K\leq N$. Each user has a cache of size $M$ files, where $0\leq M\leq N$, and the memory ratio is $M/N$. An $F$-division $(K,M,N)$ coded caching scheme consists of the following two phases.

\noindent
\textbf{Placement phase:} The server divides each file $W_n$ into $F$ equal-sized packets, i.e., $W_n=\{W_{n,j}|j\in[F]\},\ n\in[N]$. For each user $k\in[K]$, the cache content $\mathcal{Z}_k$ is selected without knowledge of future demands and has size at most $M$ files. This paper focuses on uncoded placement, in which original file packets are cached directly.

\noindent
\textbf{Delivery phase:} Each user requests one file. The demand vector is $\mathbf{d}=(d_1,\ldots,d_{K})$, where $d_k\in[N]$ means that user $k$ requests file $W_{d_k}$. Based on the demands $\mathbf{d}$ and the cache placement, the server broadcasts coded messages of total size $R_{\mathbf{d}}$ files over the shared link so that each user can recover its requested file using its cache contents. 
The maximum total size of the broadcast messages over all possible demand vectors, i.e., $R=\max_{\mathbf{d}}R_{\mathbf{d}}$, is defined as the transmission load or load for short.

\subsection{Placement Delivery Array}

Placement delivery array is proposed in \cite{YCTC} to design coded caching schemes with uncoded placement and one-shot linear delivery.

\begin{definition}[\cite{YCTC}]\label{def:pda} An $F \times K$ array $\mathbf{P}=(p_{i,j})_{i\in[F],j\in[K]}$ with entries from $\{ \ast \} \cup [S]$ is called a $(K,F,Z,S)$ placement delivery array (PDA), if it satisfies the following conditions.
	
	\textbf{C1}: Each column contains exactly $Z$ stars.
	
	\textbf{C2}: Each integer $s \in [S]$ appears at least once.
	
	\textbf{C3}: For any two distinct positions satisfying $p_{i_{1},j_{1}} = p_{i_{2},j_{2}} = s \neq \ast ,$ we have $i_{1} \neq i_{2}$, $j_{1} \neq j_{2}$, and $p_{i_{1},j_{2}} = p_{i_{2},j_{1}} = \ast$.
	
	If each integer in \([S]\) appears exactly \(g\) times in \(\mathbf P\), then \(\mathbf P\) is called a \(g\)-regular \((K,F,Z,S)\) PDA, also referred to as a \(g\)-\((K,F,Z,S)\) PDA or simply a \(g\)-PDA.
\end{definition}

A \((K,F,Z,S)\) PDA \(\mathbf P\) induces a coded caching scheme as follows:

\noindent
{\bf Placement:} Each file is divided into \(F\) packets, and user \(k\in[K]\) caches
$\mathcal Z_k=\{W_{n,j}:p_{j,k}=*,\ j\in[F],\ n\in[N]\}$.
Thus, \(p_{j,k}=*\) indicates that user \(k\) caches the \(j\)-th packet of every file. By Condition \textbf{C1}, the memory ratio is \(M/N=Z/F\).

\noindent
{\bf Delivery:} In time slot \(s\in[S]\), the server broadcasts
$
X_s=\bigoplus_{j\in[F],k\in[K],p_{j,k}=s}W_{d_k,j}$
to all users. Condition \textbf{C3} guarantees that each involved user has cached all other packets in \(X_s\) and can therefore decode its desired packet. By Condition \textbf{C2}, exactly \(S\) coded transmissions are sent. Hence, the transmission load is \(R=S/F\).

\begin{lemma}(\cite{YCTC})\label{lem:pda} A $(K,F,Z,S)$ PDA gives an $F$-division $(K,M,N)$ coded caching scheme with memory ratio \mbox{$M/N=Z/F$} and transmission load \mbox{$R=S/F$}.
\end{lemma}

\subsection{Linear Codes over Finite Fields and Rings}

Let \(\mathcal R\) denote either a finite field \(\mathbb F_q\), where \(q\) is a prime power, or the residue class ring \(\mathbb Z_q\), where \(q\geq 2\). A linear code \(\mathcal C\) of length \(n\) over \(\mathcal R\) is an \(\mathcal R\)-submodule of \(\mathcal R^n\). In particular, when \(\mathcal R\) is a finite field, \(\mathcal C\) is a linear subspace~\cite{Dougherty}. We denote the number of codewords in \(\mathcal C\) by \(|\mathcal C|\).

For any two vectors \(\boldsymbol{x},\boldsymbol{y}\in\mathcal R^n\), their Hamming distance is defined as
$d(\boldsymbol{x},\boldsymbol{y})
=|\{i\in[n]:x_i\neq y_i\}|$, i.e., the number of coordinates in which they differ. 
If \(\mathcal C\) contains at least two codewords, its minimum Hamming distance is defined as
$d_{\min}(\mathcal C)
=\min_{\boldsymbol c_1,\boldsymbol c_2\in\mathcal C,\boldsymbol c_1\neq\boldsymbol c_2}
d(\boldsymbol c_1,\boldsymbol c_2)$.

For any \(\boldsymbol{x}\in\mathcal R^n\), the set
$\boldsymbol{x}+\mathcal C
=\{\boldsymbol{x}+\boldsymbol c:\boldsymbol c\in\mathcal C\}$
is called a coset of \(\mathcal C\). The following properties will be used in our constructions:

\begin{itemize}
	\item Each coset contains exactly \(|\mathcal C|\) vectors.
	
	\item Distinct cosets are disjoint, and all distinct cosets partition \(\mathcal R^n\) into \(q^n/|\mathcal C|\) sets.
	
	\item Any two distinct vectors in the same coset have Hamming distance at least \(d_{\min}(\mathcal C)\), since adding the same vector preserves Hamming distance.
	
	\item For the zero code \(\mathcal C=\{\boldsymbol 0\}\), all its cosets are singletons. 
\end{itemize}

In particular, when \(\mathcal R=\mathbb F_q\), a linear code \(\mathcal C\) of dimension \(\delta\) contains exactly \(q^\delta\) codewords. Consequently, each coset contains \(q^\delta\) vectors, and \(\mathbb F_q^n\) is partitioned into \(q^{n-\delta}\) distinct cosets~\cite{HP}.

\section{PDA Constructions Based on Hamming Distance and Linear Codes}\label{sec:constructions}
In this paper, we propose two classes of PDA constructions that exploit Hamming distances to determine star positions and cosets of auxiliary linear codes to assign symbols to non-star entries. Positions with the same Hamming distance between their row and column labels form a layer. The first class places non-star entries only in the layer at Hamming distance \(w\), whereas the second fills all layers with Hamming distances \(r\in\{w,\ldots,m\}\) with non-star entries.

\subsection{Construction Based on a Single Hamming Distance}

\begin{construction}\label{con:one} 
	For any integers $m,w,q_1,\ldots,q_m$ satisfying $0 \leq w \leq m$, $m\geq 1$ and $q_i\geq 2$ for $i\in[m]$,
	index both the rows and columns by $\mathcal{X} = \prod_{i = 1}^{m}\left\lbrack 0:q_{i} - 1 \right\rbrack$, so $K = F = |\mathcal{X}| = \prod_{i = 1}^{m}q_{i}$. If $m+1\leq 2w$, choose a binary linear code $\mathcal{C}$ of length $w$ with the largest possible dimension $\delta$ and minimum Hamming distance at least $m-w+1$. Otherwise, let $\mathcal{C}$ be the zero code $\{\mathbf{0}_w\}$ and set $\delta=0$.
	For any $\mathbf{a},\mathbf{b} \in \mathcal{X}$, define the entries of $\mathbf{P}$ by
	\begin{equation}
		\label{eq:construction-one}
		\mathbf{P}(\mathbf{a},\mathbf{b})=
		\begin{cases}
			\ast,&d(\mathbf{a},\mathbf{b})\ne w,\\
			\bigl(\mathcal{D},\mathbf{a}_{\mathcal{D}}\cup\mathbf{b}_{\mathcal{D}},
			\boldsymbol{\xi}+\mathcal{C}\bigr),&\!d(\mathbf{a},\mathbf{b})=w.
	\end{cases}\end{equation}
	When $w>0$, $\mathcal{D} = \{ i \in \lbrack m\rbrack:a_{i} \neq b_{i}\}$ is the set of coordinates in which the row and column indices differ. If \mbox{$d(\mathbf{a},\mathbf{b}) = w$}, write $\mathcal{D} = \{ i_{1},\ldots,i_{w}\}$ in increasing order, i.e., $i_{1} < \cdots < i_{w}$.  Then ${\mathbf{a}}_{\mathcal{D}}=(a_{i_1},\ldots,a_{i_w})$ and ${\mathbf{b}}_{\mathcal{D}}=(b_{i_1},\ldots,b_{i_w})$ denote the restrictions of the two vectors to $\mathcal{D}$;  \mbox{${\mathbf{a}}_{\mathcal{D}} \cup {\mathbf{b}}_{\mathcal{D}} = ( \{ a_{i_{1}},b_{i_{1}}\},\ldots,\{ a_{i_{w}},b_{i_{w}}\} )$} is the ordered tuple of unordered symbol pairs, where the union notation is used coordinate by coordinate; $\boldsymbol{\xi}=(\xi_1,\ldots,\xi_w)$ is the binary vector used to record the ordering of the two symbols in each pair:
	\begin{equation}
		\label{eq_xi1}
		\xi_{j} = \begin{cases}
			0, & a_{i_{j}} < b_{i_{j}}, \\
			1, & a_{i_{j}} > b_{i_{j}},
		\end{cases}\quad j \in \lbrack w\rbrack,
	\end{equation}
	and $\boldsymbol{\xi} + \mathcal{C}$ denotes its coset.
	When \mbox{$w = 0$}, we define $\mathcal{D}$, ${\mathbf{a}}_{\mathcal{D}} \cup {\mathbf{b}}_{\mathcal{D}}$ and $\boldsymbol{\xi}+\mathcal{C}$ all as empty sets. 
	
	For fixed $\mathcal{D}$ and ${\mathbf{a}}_{\mathcal{D}} \cup {\mathbf{b}}_{\mathcal{D}}$, each coset $\boldsymbol{\xi}+\mathcal{C}$  groups $2^{\delta}$ binary vectors into one non-star symbol. The common values of $\mathbf{a}$ and $\mathbf{b}$ on $\mathcal{D}^{c}=[m]\setminus \mathcal{D}$ do not appear in the non-star symbol, so positions with different common values can also share that symbol. With the other parameters fixed, a larger feasible code dimension $\delta$ gives fewer distinct non-star symbols and a lower transmission load. The following example illustrates the construction.
\end{construction}

\begin{example}\label{ex:one} Let \mbox{$m = 3$}, $\left( q_{1},q_{2},q_{3} \right) = (2,2,3)$, and \mbox{$w = 2$}. Choose the binary linear code \mbox{$\mathcal{C}=\{00,11\}.$} This code has dimension $\delta=1$ and minimum distance $2$, satisfying the required distance constraint.
Using~\eqref{eq:construction-one} and relabeling all non-star symbols bijectively by $0,1,\ldots,13$, we obtain the array in \eqref{eq:pda-one}.
\begin{figure*}
	\begin{equation}
		\label{eq:pda-one}
		\mathbf{P}_{1} = \bordermatrix{
			&000  & 001  & 002  &010   &011&012&100   &101&102&110& 111  &112 \cr 
			000  &\ast & \ast & \ast & \ast & 0 & 1 & \ast & 2 & 3 & 4 & \ast & \ast \cr
			001  &\ast & \ast & \ast & 5 & \ast & 6 & 7 & \ast & 8 & \ast & 4 & \ast \cr
			002  &\ast & \ast & \ast & 9 & 10 & \ast & 11 & 12 & \ast & \ast & \ast & 4 \cr
			010  &\ast & 5 & 9 & \ast & \ast & \ast & 13 & \ast & \ast & \ast & 2 & 3 \cr
			011  & 0 & \ast & 10 & \ast & \ast & \ast & \ast & 13 & \ast & 7 & \ast & 8 \cr
			012  &1 & 6 & \ast & \ast & \ast & \ast & \ast & \ast & 13 & 11 & 12 & \ast \cr
			100  &\ast & 7 & 11 & 13 & \ast & \ast & \ast & \ast & \ast & \ast & 0 & 1 \cr
			101  & 2 & \ast & 12 & \ast & 13 & \ast & \ast & \ast & \ast & 5 & \ast & 6 \cr
			102&	3 & 8 & \ast & \ast & \ast & 13 & \ast & \ast & \ast & 9 & 10 & \ast \cr
			110&	4 & \ast & \ast & \ast & 7 & 11 & \ast & 5 & 9 & \ast & \ast & \ast \cr
			111  &	\ast & 4 & \ast & 2 & \ast & 12 & 0 & \ast & 10 & \ast & \ast & \ast \cr
			112&	\ast & \ast & 4 & 3 & 8 & \ast & 1 & 6 & \ast & \ast & \ast & \ast 
		}
	\end{equation}
\end{figure*}
For example, symbol $4$ occurs in the six positions $(00t,11t)$ and $(11t,00t)$, where $t \in \{ 0,1,2\}$. They have the same set of differing coordinates \mbox{$\mathcal{D} = \{ 1,2\}$}, unordered symbol pairs ${\mathbf{a}}_{\mathcal{D}} \cup {\mathbf{b}}_{\mathcal{D}} = \left( \{ 0,1\},\{ 0,1\} \right)$, and coset $\boldsymbol{\xi}+\mathcal{C}=\{ 00,11\}$. The cross position of the row and column at any two of the six positions is a star, since the Hamming distance between its row and column labels is not equal to $w$.
Therefore, the array in \eqref{eq:pda-one} is a $(12,12,7,14)$ PDA, which yields a $12$-user coded caching scheme with memory ratio $\frac{M}{N}=\frac{7}{12}$, subpacketization $F=12$ and load $R=\frac{7}{6}$.


\end{example}

From Construction~\ref{con:one}, we obtain the following result.

\begin{theorem}
\label{thm:one}
For any positive integers $m$, $q_{1},q_{2},\ldots,q_{m}$ with $q_{i} \geq 2$, and for any $w\in[0:m]$, if $2w\geq m+1$, let $\delta$ be the maximum possible dimension of a binary linear code of length $w$ with minimum Hamming distance at least $m - w + 1$, otherwise let $\delta=0$. Construction~\ref{con:one} gives a $\bigl(\prod_{i=1}^{m}q_i,\,\allowbreak
\prod_{i=1}^{m}q_i,\,\allowbreak
\prod_{i=1}^{m}q_i-\sum_{\mathcal D\subseteq\lbrack m\rbrack,\,|\mathcal D|=w}\prod_{i\in\mathcal D}(q_i-1),\,\allowbreak
2^{w-\delta}\sum_{\mathcal D\subseteq\lbrack m\rbrack,\,|\mathcal D|=w}\prod_{i\in\mathcal D}\binom{q_i}{2}\bigr)$~PDA, which yields a $(K,M,N)$ coded caching scheme with the following parameters
$$\textstyle K=F=\prod_{i = 1}^{m}q_{i},\ \
\frac{M}{N} = 1 - \frac{\sum_{\mathcal D\subseteq\lbrack m\rbrack,\,|\mathcal D|=w}{\prod_{i \in \mathcal{D}}(q_i-1)}}{\prod_{i = 1}^{m}q_{i}},\ \
R = \frac{2^{w - \delta}\sum_{\mathcal D\subseteq\lbrack m\rbrack,\,|\mathcal D|=w}{\prod_{i \in \mathcal{D}}\binom{q_{i}}{2}}}{\prod_{i = 1}^{m}q_{i}}.$$
\end{theorem}

The proof of Theorem \ref{thm:one} is given in Appendix~\ref{app:one}. 
When all coordinate alphabets have the same size, i.e., $q_{1}=\ldots=q_{m}=q$, we obtain the following result.

\begin{theorem}
\label{cor:one}
For any integers \(m,q,w\) satisfying \(0\leq w\leq m\), \(m\geq1\), and \(q\geq2\), let \(\delta\) be the maximum possible dimension of a binary linear code of length \(w\) with minimum Hamming distance at least \(m-w+1\) if \(2w\geq m+1\), and set \(\delta=0\) otherwise. 
Let $\mathcal{C}'$ be a $\mathbb{Z}_q$-linear code of length $m-w$ with the maximum possible cardinality $\nu=|\mathcal{C}'|$ subject to a minimum Hamming distance of at least \(w+1\) if $2w<=m-1$, otherwise let $\mathcal{C}'$ be the zero code $\{{\bf 0}_{m-w}\}$ and set $\nu=1$.
Then, there exists a $g=\max\{2^{\delta}q^{m-w},2^w\nu\}$-$\big(
q^m,\,
q^m,\,
q^m-\binom{m}{w}(q-1)^w,\,
\frac{\binom{m}{w}(q-1)^wq^m}{g}
\big)$ 
PDA, which leads to a $(K,M,N)$ coded caching scheme with the number of users $K=q^m$, the memory ratio $\frac{M}{N}=1-	\frac{\binom{m}{w}(q-1)^w}{q^m}$, the subpacketization $F=q^m$,
and the transmission load 
$R=
\frac{\binom{m}{w}(q-1)^w}
{g}$.
\end{theorem}

\begin{proof} 
When \mbox{$q_{1} = \cdots = q_{m} = q$}, we have
\begin{equation*}\begin{aligned}\textstyle \sum_{\mathcal D\subseteq\lbrack m\rbrack,\,|\mathcal D|=w}{\prod_{i \in \mathcal{D}}(q_i-1)} &\textstyle= \binom{m}{w}(q - 1)^{w}, \\ \textstyle \sum_{\mathcal D\subseteq\lbrack m\rbrack,\,|\mathcal D|=w}{\prod_{i \in \mathcal{D}}\binom{q_{i}}{2}} &\textstyle= \binom{m}{w}\binom{q}{2}^{w}.\end{aligned}\end{equation*}
Substituting them into Theorem~\ref{thm:one} gives a $2^{\delta}q^{m-w}$-$\big(
q^m,\,
q^m,\,
q^m-\binom{m}{w}(q-1)^w,\,
\frac{\binom{m}{w}(q-1)^wq^m}{2^{\delta}q^{m-w}}
\big)$ PDA. For a fixed non-star label, the coset contains $2^{\delta}$ binary vectors, and there are $q^{m - w}$ choices for the common coordinates in $\mathcal{D}^{c}=[m]\setminus \mathcal{D}$. Every non-star symbol therefore appears exactly $2^{\delta}q^{m - w}$ times, proving regularity.

On the other hand, for the same star placement we can also construct a square PDA as follows:
\begin{construction}
	\label{constr2}
	Let \(m,q,w\) be integers satisfying
	$0\leq w\leq m$, $m\geq 1$ and $q\geq 2$.
	If $m\geq 2w+1$, choose a $\mathbb{Z}_q$-linear code $\mathcal{C}'$ of length $m-w$ with the largest possible cardinality $\nu=|\mathcal{C}'|$ subject to a minimum Hamming distance of at least \(w+1\). Otherwise, let $\mathcal{C}'$ be the zero code $\{\mathbf{0}_{m-w}\}$ and set $\nu=1$.
	For any
	$\mathbf{a},\mathbf{b}\in[0:q-1]^m$,
	define
	\begin{equation}
		\label{eq:construction2}
		\mathbf{P}(\mathbf{a},\mathbf{b})
		=
		\begin{cases}
			*,&
			d(\mathbf{a},\mathbf{b})\neq w,\\[1ex]
			\left(
			\mathcal{D},\,
			\mathbf{a}_{\mathcal{D}}\cup\mathbf{b}_{\mathcal{D}},\,
			\mathbf{a}_{\mathcal{D}^{c}}+\mathcal{C}'
			\right),&
			d(\mathbf{a},\mathbf{b})=w.
		\end{cases}		
	\end{equation}
	If $w>0$, $\mathcal{D}= \{ i \in \lbrack m\rbrack:a_{i} \neq b_{i}\}$ denotes the set of coordinates in which $\mathbf{a}$ and $\mathbf{b}$ differ; $\mathcal{D}^{c}=[m]\setminus\mathcal{D}$ denotes the complement of 
	$\mathcal{D}$ with respect to the universal set 
	$[m]$. If $w=0$, $\mathcal{D}$ and $\mathbf{a}_{\mathcal{D}}\cup\mathbf{b}_{\mathcal{D}}$ are defined as empty sets, $\mathcal{D}^{c}=[m]$, and $\mathbf{a}_{\mathcal{D}^{c}}+\mathcal{C}'=[0:q-1]^m$. 
\end{construction}
The following result is obtained via Construction~\ref{constr2}, with its proof provided in Appendix~\ref{app_lemm1}.
\begin{lemma}
	\label{lemm1}
	The array generated by Construction \ref{constr2} is a $2^w\nu$-$\big(q^m,\,q^m,\,q^m-\binom{m}{w}(q-1)^w,\,\frac{q^m\binom{m}{w}(q-1)^w}{2^w\nu}
	\big)$ PDA.	
\end{lemma}
For given $m, q, w$, we can choose between Construction \ref{con:one} and Construction \ref{constr2} the one that yields the larger multicast gain, thereby obtaining a 
$g=\max\{2^{\delta}q^{m-w},2^w\nu\}$-$\big(
q^m,\,
q^m,\,
q^m-\binom{m}{w}(q-1)^w,\,
\frac{\binom{m}{w}(q-1)^wq^m}{g}
\big)$ 
PDA, which leads to a $(K,M,N)$ coded caching scheme with the number of users $K=q^m$, the memory ratio $\frac{M}{N}=1-	\frac{\binom{m}{w}(q-1)^w}{q^m}$, the subpacketization $F=q^m$,
and the transmission load 
$R=
\frac{\binom{m}{w}(q-1)^w}
{g}$.
\end{proof}

\subsection{Construction Based on Consecutive Hamming Distances}
Constructions 1 and 2 generate square PDAs by placing non-star entries in a single Hamming-distance layer, which generally leads to relatively high memory ratios. To achieve greater flexibility in memory ratios, we introduce a new construction in this section. Specifically, for any \(w\in[m]\), we assign non-star entries to all Hamming-distance layers \(r\in[w:m]\), while placing stars at positions where the Hamming distance is less than \(w\).

\begin{construction}\label{con:two} Let $m, w$ be positive integers with $w\leq m$, let $q_{1},q_{2},\ldots,q_{m}$ be integers satisfying $q_{i} \geq 2$. Index both the rows and columns by $\mathcal{X} = \prod_{i = 1}^{m}\left\lbrack 0:q_{i} - 1 \right\rbrack$, so \mbox{$K = F = \prod_{i = 1}^{m}q_{i}$}. For each $r\in\lbrack w:m\rbrack$, let $\mathcal{C}_r$ be a binary linear code of length $r$ with maximum possible dimension $\delta_r$ and minimum Hamming distance at least $r-w+1$.
For any $\mathbf{a},\mathbf{b} \in \mathcal{X}$, let \mbox{$r = d(\mathbf{a},\mathbf{b})$} and define
{\setlength{\arraycolsep}{2pt}\begin{equation}\label{eq:construction-two}\mathbf{P}(\mathbf{a},\mathbf{b})=
		\begin{cases}
			\ast,&r<w,\\
			\bigl(\mathcal{D},\mathbf{a}_{\mathcal{D}}\cup\mathbf{b}_{\mathcal{D}},
			\mathbf{a}_{\mathcal{D}^c},\boldsymbol{\xi}+\mathcal{C}_r\bigr),&r\ge w.
\end{cases}\end{equation}}
Here $\mathcal{D}= \{ i \in \lbrack m\rbrack:a_{i} \neq b_{i}\}$ is the set of coordinates where the two indices differ, and \mbox{$|\mathcal{D}| = r$}; let $\mathcal{D} = \{ i_{1},i_{2},\ldots,i_{r}\}$ in increasing order, i.e., $i_{1} < i_{2} < \cdots < i_{r}$; \mbox{${\mathbf{a}}_{\mathcal{D}} \cup {\mathbf{b}}_{\mathcal{D}} = ( \{ a_{i_{1}},b_{i_{1}}\},\ldots,\{ a_{i_{r}},b_{i_{r}}\} )$} is an ordered tuple of unordered pairs, where ${\mathbf{a}}_{\mathcal{D}}$ and ${\mathbf{b}}_{\mathcal{D}}$ are the restrictions to $\mathcal{D}$; $\boldsymbol{\xi}=(\xi_1,\ldots,\xi_r)$ is the binary vector used to record the ordering of the two symbols in each pair:
\begin{equation}
	\label{eq_xi2}
	\xi_{j} = \begin{cases}
		0, & a_{i_{j}} < b_{i_{j}}, \\
		1, & a_{i_{j}} > b_{i_{j}},
	\end{cases}\quad j \in \lbrack r\rbrack.
\end{equation} 
$\mathbf{a}_{\mathcal{D}^c}$ denotes the restriction of $\mathbf{a}$ to $\mathcal{D}^c=[m]\setminus \mathcal{D}$, which records the common values between $\mathbf{a}$ and $\mathbf{b}$. If $r=m$, then $\mathcal{D}^c=\emptyset$ and $\mathbf{a}_{\mathcal{D}^c}$ is defined as $\emptyset$. 




%

For fixed $\mathcal{D}$, $\mathbf{a}_{\mathcal{D}}\cup\mathbf{b}_{\mathcal{D}}$, and $\mathbf{a}_{\mathcal{D}^c}$, each coset $\boldsymbol{\xi}+\mathcal{C}_r$  groups $2^{\delta_r}$ binary vectors into one non-star symbol. Therefore, larger feasible code dimensions $\delta_r$ ($r\in[w:m]$) reduce the number of non-star symbols and the transmission load when the other parameters are fixed. The following example illustrates Construction~\ref{con:two}.
\end{construction}

\begin{example}\label{ex:two} Let $m = 3$, $\left( q_{1},q_{2},q_{3} \right) = (2,2,3)$, and $w = 2$. For the layer at distance $2$, we use the full-space code $\mathcal{C}_{2} = \mathbb{F}_{2}^{2} = \{ 00,01,10,11\}$ with dimension $\delta_{2}=2$. For the layer at distance $3$, we use the single-parity-check code $\mathcal{C}_{3} = \{ 000,011,101,110\}$ with dimension $\delta_{3} = 2$. Both $\mathcal{C}_{2}$ and $\mathcal{C}_{3}$ satisfy the respective distance bounds. Using~\eqref{eq:construction-two} and relabeling all non-star symbols bijectively by $0,1,\ldots,20$, we obtain the array in \eqref{eq:pda-two}. For example, the subarray containing symbol $4$ is 
\begin{equation*}
	\bordermatrix{
		&000&010  &100   &110\cr
		000&\ast & \ast & \ast & 4 \cr
		010&\ast & \ast & 4 & \ast \cr
		100&\ast & 4    & \ast & \ast \cr
		110&4    & \ast & \ast & \ast},
\end{equation*} 
where the four non-star positions lie in the layer with Hamming distance $r=2$ and have the same $\mathcal{D}=\{1,2\}$,  $\mathbf{a}_{\mathcal{D}}\cup\mathbf{b}_{\mathcal{D}}=(\{0,1\},\{0,1\})$, and $\mathbf{a}_{\mathcal{D}^c}=(0)$. The binary vectors of the four non-star pisitions from left to right are $\boldsymbol{\xi}_1=(1,1)$, $\boldsymbol{\xi}_2=(1,0)$,
$\boldsymbol{\xi}_3=(0,1)$, and
$\boldsymbol{\xi}_4=(0,0)$ respectively. They are grouped into the same coset $\boldsymbol{\xi}_i+\mathcal{C}_2=\mathcal{C}_2$ for $i\in[4]$. For another example, the subbarray containing symbol $9$ is
\begin{equation*}
	\bordermatrix{
		&000&011  &101   &110\cr
		001&\ast & \ast & \ast & 9 \cr
		010&\ast & \ast & 9 & \ast \cr
		100&\ast & 9    & \ast & \ast \cr
		111&9    & \ast & \ast & \ast},
\end{equation*} 
where the four non-star positions lie in the layer with Hamming distance $r=3$ and have the same $\mathcal{D}=\{1,2,3\}$,  $\mathbf{a}_{\mathcal{D}}\cup\mathbf{b}_{\mathcal{D}}=(\{0,1\},\{0,1\},\{0,1\})$, and $\mathbf{a}_{\mathcal{D}^c}=\emptyset$. The binary vectors of the four non-star pisitions from left to right are $\boldsymbol{\xi}_1=(1,1,1)$, $\boldsymbol{\xi}_2=(1,0,0)$,
$\boldsymbol{\xi}_3=(0,1,0)$, and
$\boldsymbol{\xi}_4=(0,0,1)$ respectively. They are grouped into the same coset $\boldsymbol{\xi}_i+\mathcal{C}_3=\{001,010,100,111\}$ for $i\in[4]$. The array in \eqref{eq:pda-two} is a $(12,12,5,21)$ PDA, which yields a coded caching scheme with $K=F=12$, memory ratio $M/N=5/12$, and load $R=21/12=7/4$.
\begin{figure*}
	\begin{equation}
		\label{eq:pda-two}
		\mathbf{P}_{2} = \bordermatrix{
			&000  & 001  & 002  &010   &011&012&100   &101&102&110& 111  &112 \cr 
			000 &			\ast & \ast & \ast & \ast & 0 & 1 & \ast & 2 & 3 & 4 & 5 & 6 \cr
			001  &			\ast & \ast & \ast & 0 & \ast & 7 & 2 & \ast & 8 & 9 & 10 & 11 \cr
			002  &			\ast & \ast & \ast & 1 & 7 & \ast & 3 & 8 & \ast & 12 & 13 & 14 \cr
			010   &			\ast & 0 & 1 & \ast & \ast & \ast & 4 & 9 & 12 & \ast & 15 & 16 \cr
			011&			0 & \ast & 7 & \ast & \ast & \ast & 5 & 10 & 13 & 15 & \ast & 17 \cr
			012&			1 & 7 & \ast & \ast & \ast & \ast & 6 & 11 & 14 & 16 & 17 & \ast \cr
			100   &			\ast & 2 & 3 & 4 & 9 & 12 & \ast & \ast & \ast & \ast & 18 & 19 \cr
			101&			2 & \ast & 8 & 5 & 10 & 13 & \ast & \ast & \ast & 18 & \ast & 20 \cr
			102&			3 & 8 & \ast & 6 & 11 & 14 & \ast & \ast & \ast & 19 & 20 & \ast \cr
			110&			4 & 5 & 6 & \ast & 15 & 16 & \ast & 18 & 19 & \ast & \ast & \ast \cr
			111  &			9 & 10 & 11 & 15 & \ast & 17 & 18 & \ast & 20 & \ast & \ast & \ast \cr
			112&			12 & 13 & 14 & 16 & 17 & \ast & 19 & 20 & \ast & \ast & \ast & \ast }
	\end{equation}
\end{figure*}
\end{example}

In general, the following theorem can be obtained from Construction~\ref{con:two}.

\begin{theorem}
\label{thm:two}
For any positive integers $m,w$ with $w\leq m$, let $q_{1},q_{2},\ldots,q_{m}$ be integers satisfying $q_{i} \geq 2$. For each $r \in \lbrack w:m\rbrack$, let $\delta_r$ denote the maximum possible dimension of a binary linear code of length $r$ with minimum Hamming distance at least $r - w + 1$. Construction~\ref{con:two} gives a $\bigl(\prod_{i=1}^{m}q_i,\,\allowbreak
\prod_{i=1}^{m}q_i,\,\allowbreak
\prod_{i=1}^{m}q_i-\sum_{r=w}^{m}\sum_{\mathcal D\subseteq\lbrack m\rbrack,\,|\mathcal D|=r}\prod_{i\in\mathcal D}(q_i-1),\,\allowbreak
\prod_{i=1}^{m}q_i\sum_{r=w}^{m}\frac{1}{2^{\delta_r}}\sum_{\mathcal D\subseteq\lbrack m\rbrack,\,|\mathcal D|=r}$ $\prod_{i\in\mathcal D}(q_i-1)\bigr)$~PDA, which yields a $(K,M,N)$ coded caching scheme with the following parameters
$$\textstyle K=F=\prod_{i = 1}^{m}q_{i},\ \ \textstyle \frac{M}{N} = 1 - \frac{\sum_{r = w}^{m}{\sum_{\mathcal D\subseteq\lbrack m\rbrack,\,|\mathcal D|=r}{\prod_{i \in \mathcal{D}}(q_i-1)}}}{\prod_{i = 1}^{m}q_{i}},\ \
\textstyle R = \sum_{r = w}^{m}\frac{1}{2^{\delta_r}}\sum_{\mathcal D\subseteq\lbrack m\rbrack,\,|\mathcal D|=r}{\prod_{i \in \mathcal{D}}(q_i-1)}.$$

\end{theorem}
The proof of Theorem \ref{thm:two} is given in Appendix~\ref{app:two}. When all coordinate alphabets have the same size, i.e., $q_1=\ldots=q_m=q$, Theorem~\ref{thm:two} yields the following corollary.

\begin{corollary}
\label{cor:two}
For any positive integers $m,w,q$ with $w\leq m$ and $q\geq 2$, for each $r \in \lbrack w:m\rbrack$, let $\delta_r$ denote the maximum possible dimension of a binary linear code of length $r$ with minimum Hamming distance at least $r - w + 1$.	
There exists a $\bigl(q^m,q^m,\,\allowbreak q^m-\sum_{r=w}^{m}\binom{m}{r}(q-1)^r,\,\allowbreak q^m\sum_{r=w}^{m}\frac{\binom{m}{r}(q-1)^r}{2^{\delta_r}}\bigr)$~PDA, which yields a $(K,M,N)$ coded caching scheme with
$K = F = q^{m}$, memory ratio $\frac{M}{N} = 1 - \frac{1}{q^{m}}\sum_{r = w}^{m}\binom{m}{r}(q - 1)^{r}$, and load $R = \sum_{r = w}^{m}\frac{1}{2^{\delta_r}}\binom{m}{r}(q - 1)^{r}$.
\end{corollary}

\section{Performance Analysis}
\label{sec:performance}

In this section, we compare the proposed schemes with existing coded caching schemes in terms of transmission load and subpacketization. We first conduct theoretical comparisons with existing schemes, whenever possible, under the same number of users and memory ratio. We then present numerical comparisons.

\subsection{Theoretical Comparisons}

\subsubsection{Comparison of the schemes in Theorem~\ref{cor:one} and Corollary~\ref{cor:two} with the MN scheme in~\cite{MN}}
Both proposed schemes have \(K=q^m\) users and subpacketization \(F=K\), but they employ different cache placements and hence achieve different memory ratios. Specifically, let
$A_1=\binom{m}{w}(q-1)^w$ and
$A_2=\sum_{r=w}^{m}\binom{m}{r}(q-1)^r$.
The memory ratios of the schemes in Theorem 2 and Corollary 1 are \(M/N=1-A_1/K\) and \(M/N=1-A_2/K\), respectively.

For Theorem~\ref{cor:one}, to achieve the same memory ratio, the MN scheme requires \(t=K-A_1\), and hence has transmission load $R_{\mathrm{MN}}
=\frac{A_1}{K-A_1+1}$ and subpacketization
$F_{\mathrm{MN}}=\binom{K}{A_1}$.
In contrast, Theorem 2 achieves
$R_{\mathrm{Th2}}=\frac{A_1}{g}$ and 
$F_{\mathrm{Th2}}=K$,
where \(g=\max\{2^\delta q^{m-w},2^w\nu\}\). The comparison can be summarized as follows:
\begin{itemize}
\item If \(w=0\), then \(A_1=1\), \(\delta=0\), and \(\nu=q^m\). Hence, \(R_{\mathrm{Th2}}=R_{\mathrm{MN}}=1/K\) and \(F_{\mathrm{Th2}}=F_{\mathrm{MN}}=K\).

\item If \(q=2\) and \(w=m\), then \(A_1=1\), \(\delta=m\), and \(\nu=1\). Thus, \(R_{\mathrm{Th2}}=R_{\mathrm{MN}}=1/K\) and \(F_{\mathrm{Th2}}=F_{\mathrm{MN}}=K\).

\item If \(m=w=1\) and \(q\geq2\), then \(A_1=q-1=K-1\), \(\delta=1\), and \(\nu=1\). It follows that \(R_{\mathrm{Th2}}=R_{\mathrm{MN}}=(K-1)/2\) and \(F_{\mathrm{Th2}}=F_{\mathrm{MN}}=K\).

\item For all the remaining parameter choices, the MN scheme achieves a strictly lower transmission load, whereas Theorem 2 requires only \(K\) subpackets, compared with \(\binom{K}{A_1}\) subpackets for the MN scheme.
\end{itemize}

For Corollary~\ref{cor:two}, the MN scheme requires \(t=K-A_2\) to achieve the same memory ratio, yielding
$R_{\mathrm{MN}}=\frac{A_2}{K-A_2+1}$ and
$F_{\mathrm{MN}}=\binom{K}{A_2}$.
Corollary~\ref{cor:two} achieves
$R_{\mathrm{Co1}}
=\sum_{r=w}^{m}\binom{m}{r}
\frac{(q-1)^r}{2^{\delta_r}}$. The load and subpacketization comparisons between the two schemes are detailed as follows:
\begin{itemize}
\item When $w=1$, we have $\delta_r=1$ for all 
$r\in[1:m]$ and $A_2=K-1$. It follows that $R_{\mathrm{Co1}}=R_{\mathrm{MN}}
=\frac{K-1}{2}$, $F_{\mathrm{Co1}}=F_{\mathrm{MN}}=K$.

\item When $q=2$ and $w=m$, we have $A_2=1$ and $\delta_m=m$. Consequently,
$R_{\mathrm{Co1}}=R_{\mathrm{MN}}
=\frac{1}{K}$, $F_{\mathrm{Co1}}=F_{\mathrm{MN}}=K$.

\item In all other cases, the MN scheme achieves a 
strictly lower transmission load, whereas 
Corollary~\ref{cor:two} substantially reduces the 
subpacketization, requiring only $K$ packets instead 
of $\binom{K}{A_2}$ packets in the MN scheme.
\end{itemize}

\subsubsection{Comparison of the scheme in Theorem~\ref{cor:one} with the WCCLS scheme in~\cite{WCCLS}}
For both schemes, the number of users and subpacketization 
are $K=F=q^m$, and the memory ratio is
$\frac{M}{N}
=1-\frac{\binom{m}{w}(q-1)^w}{q^m}$.
Their transmission loads are given by
\[
R_{\mathrm{WCCLS}}
=\frac{\binom{m}{w}(q-1)^w}{q^{m-w}},
\
R_{\mathrm{Th2}}
=\frac{\binom{m}{w}(q-1)^w}
{\max\{2^\delta q^{m-w},2^w\nu\}},
\]
respectively, where $\delta$ and $\nu$ are
defined in Theorem~\ref{cor:one}. Consequently,
\[
\frac{R_{\mathrm{Th2}}}{R_{\mathrm{WCCLS}}}
=\min\left\{\frac{1}{2^\delta},
\frac{q^{m-w}}{2^w\nu}\right\}.
\]
In particular, the following two cases can be distinguished.
\begin{itemize}
\item If $2w\leq m$, then $\delta=0$. Moreover, when $2w<m$, by the
Singleton bound~\cite{Singleton1964}, $\nu=|\mathcal{C}'|\leq q^{m-2w}$, since
$\mathcal{C}'$ is a $\mathbb Z_q$-linear code of length $m-w$ with minimum Hamming distance at least $w+1$. Consequently,
$2^w\nu\leq 2^w q^{m-2w}
\leq q^{m-w}$,
where the last inequality follows from $q\geq 2$. When $2w=m$, we have $\nu=1$, then $2^w\nu=2^w\leq q^w=q^{m-w}$.
Hence, $R_{\mathrm{Th2}}=R_{\mathrm{WCCLS}}$.

\item If $2w\geq m+1$, then
$\delta\geq 1$, since a binary repetition code of 
length $w$ satisfies the required minimum distance 
constraint. Consequently,
$	\frac{R_{\mathrm{Th2}}}{R_{\mathrm{WCCLS}}}
\leq\frac{1}{2^\delta}$,
implying that the scheme in Theorem~\ref{cor:one} 
reduces the transmission load by at least $50\%$.
\end{itemize}
Therefore, compared with the WCCLS scheme, the scheme in Theorem~\ref{cor:one} achieves the same transmission load when $2w\leq m$, and reduces the transmission load by at least $50\%$ when $2w\geq m+1$, while maintaining the same subpacketization and memory ratio.

\subsubsection{Comparison of the scheme in Theorem~\ref{cor:one} with the CWWC scheme in~\cite{CWWC}}
For an odd integer \(q\geq3\), setting \(v=q^m\) and \(n=m\) in the CWWC scheme in~\cite{CWWC} yields
\[
K=F=q^m,\quad
\frac{M}{N}=1-\frac{(q-1)^m}{q^m},\quad
R_{\mathrm{CWWC}}=\frac{(q-1)^m}{2^m}.
\]
On the other hand, the scheme in Theorem~\ref{cor:one} has
\[
K=F=q^m,\quad
\frac{M}{N}=1-\frac{A}{q^m},\quad
R_{\mathrm{Th2}}=\frac{A}{g},
\]
where
$A=\binom{m}{w}(q-1)^w$,
$g=\max\{2^\delta q^{m-w},2^w\nu\}$.
To compare the two schemes under the same number of users and memory ratio, we require
$\binom{m}{w}=(q-1)^{m-w}$.
Under this condition, both schemes have the same subpacketization \(F=K=q^m\), and their transmission-load ratio is
\[
\frac{R_{\mathrm{Th2}}}{R_{\mathrm{CWWC}}}
=\frac{2^m}{\max\{2^\delta q^{m-w},2^w\nu\}}.
\]
By the definitions of \(\delta\) and \(\nu\) in Theorem~\ref{cor:one}, this ratio can be further expressed as
\[
\frac{R_{\mathrm{Th2}}}{R_{\mathrm{CWWC}}}
=
\begin{cases}
\displaystyle
\min\left\{
\frac{2^m}{q^{m-w}},
\frac{2^{m-w}}{\nu}
\right\},
& 2w\leq m-1,\\[3mm]
\displaystyle
\left(\frac{4}{q}\right)^w,
& 2w=m,\\[3mm]
\displaystyle
\min\left\{
\frac{2^{m-\delta}}{q^{m-w}},
2^{m-w}
\right\},
& 2w\geq m+1.
\end{cases}
\]
Consequently, Theorem~\ref{cor:one} achieves a lower transmission load if this ratio is less than one, whereas the CWWC scheme achieves a lower load if it is greater than one. In particular, the following two cases can be identified.

\begin{itemize}
\item If \(w=m\), then \(\delta=m\) and \(\nu=1\). Hence, the two schemes achieve exactly the same performance.

\item If \(w=m-1\), the condition for achieving the same memory ratio becomes \(q=m+1\). Since \(q\) is odd, \(m\) must be even. For \(m\geq4\), we have \(\delta=m-2\) and \(\nu=1\), yielding
$\frac{R_{\mathrm{Th2}}}{R_{\mathrm{CWWC}}}
=\frac{4}{m+1}<1$.
Thus, for even \(m\geq4\), the scheme in Theorem~\ref{cor:one} achieves a strictly lower transmission load than the CWWC scheme.
\end{itemize}

\subsection{Numerical Comparisons}
\subsubsection{Comparison with existing linear-subpacketization schemes}
We first compare the proposed schemes in Theorem \ref{cor:one} and Corollary \ref{cor:two} with existing schemes that have linear subpacketization in the number of users, focusing on the tradeoff between transmission load and memory ratio. Specifically, the proposed schemes and most of the considered existing schemes have a subpacketization of \(F=K\), whereas the ZCW schemes~\cite{ZCW} achieve $F\leq K$, and the YCWC scheme~\cite{YCWC} has \(F=gK\) where \(g\) denotes its coded multicast gain.

\begin{figure}
\centering
\includegraphics[width=4in]{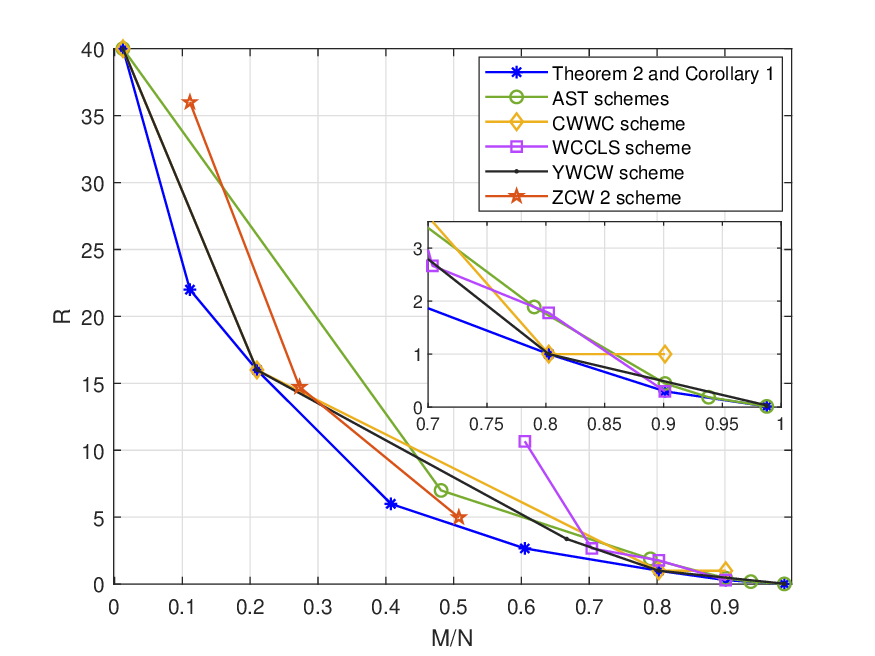}\\
\caption{Transmission load versus memory ratio for $K=81$.}\label{Fig_K_81}
\end{figure}

For \(K=81\), an odd number, Fig. \ref{Fig_K_81} shows that the proposed schemes achieve lower transmission loads than the existing AST~\cite{AST}, ZCW 2~\cite{ZCW}, WCCLS~\cite{WCCLS}, CWWC~\cite{CWWC}, and YCWC~\cite{YCWC} schemes at many comparable memory ratios, demonstrating their competitive load performance with a subpacketization level of \(F=K\).

\begin{figure}[!htbp]
\centering
\includegraphics[width=4in]{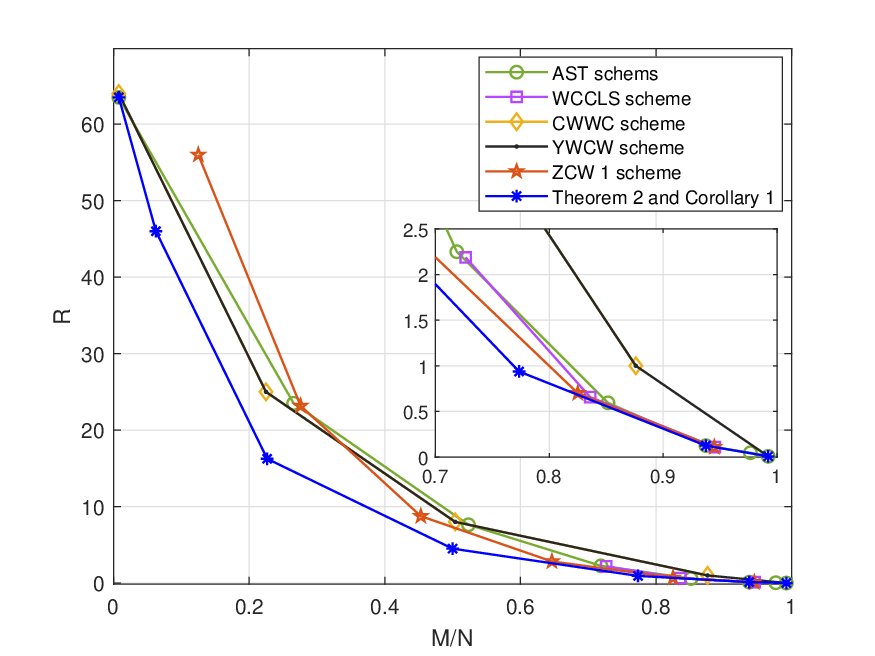}\\
\caption{Transmission load versus memory ratio for $K=128$.}\label{Fig_K_128}
\end{figure}

For \(K=128\), an even number, we also include the CWWC and YCWC schemes by introducing one virtual user, resulting in an effective user parameter of \(v=129\). As shown in Fig. \ref{Fig_K_128}, the proposed schemes also achieve a favorable load-memory tradeoff, outperforming the existing AST~\cite{AST}, ZCW 1~\cite{ZCW}, WCCLS~\cite{WCCLS}, CWWC~\cite{CWWC}, and YCWC~\cite{YCWC} schemes over a wide range of memory ratios. 

Table~\ref{tab:nhsdp} presents numerical comparisons with existing schemes under the same or comparable memory ratios. The results show that the proposed schemes can achieve lower transmission loads than existing WCCLS, ZCW, ASK, AST, CWWC, and YCWC schemes. Moreover, all proposed schemes maintain linear subpacketization \(F=K\), which is smaller than that of some existing schemes. For example, compared with ASK 2, the proposed scheme reduces \(F\) from \(20\) to \(16\) while lowering the transmission load from \(0.8\) to \(0.5\). More notably, compared with YCWC, it reduces \(F\) from \(1400\) to \(175\) and the transmission load from \(8\) to \(5.9429\). These comparisons demonstrate the advantages of the proposed schemes in balancing transmission load and subpacketization.
\begin{table}
\centering
\fontsize{8}{9.5}\selectfont
\setlength{\tabcolsep}{1.5pt}
\renewcommand{\arraystretch}{1.00}
\caption{Numerical comparison with existing schemes having linear subpacketization}\label{tab:nhsdp}
\begin{tabular}{|c|c|c|c|c|c|}
	\hline
	$K$ & $M/N$ & Scheme & Parameters & $R$ & $F$ \\
	\hline
	
	27 & 0.5556 & WCCLS~\cite{WCCLS} & $(m,q,w)=(3,3,2)$ & 4 & 27 \\
	27 & 0.5556 & Theorem~\ref{cor:one} & $(m,q,w,\delta,\nu)=(3,3,2,1,1)$ & 2 & 27 \\
	\hline

	256 & 0.9686 & ZCW 1~\cite{ZCW} & $(m,w)=(8,7)$ & 0.0627 & 255 \\
	256 & 0.9688 & Theorem~\ref{cor:one} & $(m,q,w,\delta,\nu)=(8,2,1,0,64)$ & 0.0625 & 256 \\
	\hline	
	
	729 &  0.0769 & ZCW 2~\cite{ZCW} & $(m,w)=(6,1)$ & 336.4615 & 13 \\
	729 &  0.0727 & Theorem~\ref{cor:one} & $(m,q,w,\delta,\nu)=(2,27,2,2,1)$ & 169	 & 729 \\
	\hline		
	
	273 & 0.9377 & ASK 1~\cite{ASK} & $q=16$ & 1 & 273 \\
	273 & 0.9267 & Theorem~\ref{thm:one} & $(m,q_1,q_2,q_3,w,\delta)=(3,3,7,13,1,0)$ & 0.7473 & 273 \\
	\hline
	
	16 & 0.75 & ASK 2~\cite{ASK} & $q=4$ & 0.8 & 20 \\
	16 & 0.75 & Theorem~\ref{cor:one} & $(m,q,w,\delta,\nu)=(4,2,3,2,1)$ & 0.5 & 16 \\
	\hline
	
	
	64 & 0.7656 & AST 1~\cite{AST} & $g=8$ & 1.875 & 64 \\
	64 & 0.7656 & Theorem~\ref{cor:one} & $(m,q,w,\delta,\nu)=(6,2,2,0,2)$ & 0.9375 & 64 \\
	\hline
	
	32 & 0.6875 & AST 2~\cite{AST} & $g=4$ & 1.25 & 32 \\
	32 & 0.6875 & Theorem~\ref{thm:one} & $(m,q_1,\ldots,q_4,w,\delta)=(4,2,2,2,4,3,2)$ & 1.1875 & 32 \\
	\hline
	
	80 & 0.3875 & AST 3~\cite{AST} & $(g,d)=(20,2)$ & 12.25 & 80 \\
	80 & 0.3875 & Theorem~\ref{thm:two} & \tabincell{c}{$(m,q_1,\ldots,q_5,w)=$\ $(5,2,2,2,2,5,3)$, $(\delta_3,\delta_4,\delta_5)=(3,3,2)$} & 6.625 & 80 \\
	\hline
	
	96 & 0.375 & AST 4~\cite{AST} & $(g,d)=(24,2)$ & 15 & 96 \\
	96 & 0.375 & Theorem~\ref{thm:two} & \tabincell{c}{$(m,q_1,\ldots,q_5,w)=(5,2,2,2,2,6,3)$, $(\delta_3,\delta_4,\delta_5)=(3,3,2)$} & 8.125 & 96 \\
	\hline
	
	72 & 0.4861 & AST 5~\cite{AST} & $(g,d)=(24,3)$ & 6.1667 & 72 \\
	72 & 0.4861 & Theorem~\ref{thm:two} & \tabincell{c}{$(m,q_1,\ldots,q_4,w)=(4,2,2,3,6,3)$, $(\delta_3,\delta_4)=(3,3)$} & 4.625 & 72 \\
	\hline
	
	864 & 0.7130 & AST 6~\cite{AST} & $n=6$ & 20.6667 & 864 \\
	864 & 0.7130 & Theorem~\ref{thm:one} & $(m,q_1,\ldots,q_4,w,\delta)=(4,3,3,3,32,4,4)$ & 15.5 & 864 \\
	\hline
	
	16 & 0.4375 & AST 7~\cite{AST} & $(b,g,d)=(8,2,2)$ & 4.5 & 16 \\
	16 & 0.4375 & Corollary~\ref{cor:two} & $(m,q,w,\delta_2)=(2,4,2,2)$ & 2.25 & 16\\
	\hline
	
	512 & 0.2539 & AST 8~\cite{AST} & $(k,r)=(128,2)$ & 95.5 & 512 \\
	512 & 0.2539 & Corollary~\ref{cor:two} & $(m,q,w,\delta_4,\ldots,\delta_9)=(9,2,4,4,4,3,3,2,2)$ & 33.25 & 512 \\
	\hline
	
	45 & 0.6444 & CWWC~\cite{CWWC} & $(v,n)=(45,2)$ & 4 & 45 \\
	45 & 0.6444 & Theorem~\ref{thm:two} & $(m,q_1,q_2,q_3,w,\delta_3)=(3,3,3,5,3,3)$ & 2 & 45 \\
	\hline	
	
	175 & 0.6343 & YCWC~\cite{YCWC} & $(v,n)=(175,3)$ & 8 & 1400 \\
	175 & 0.6343 & Theorem~\ref{thm:one} & $(m,q_1,q_2,q_3,w,\delta)=(3,5,5,7,2,1)$ & 5.9429 & 175 \\
	\hline
	
	
\end{tabular}
\end{table}

\subsubsection{Comparison with other existing coded caching schemes}

\begin{figure}[!htbp]
\centering
\includegraphics[width=4in]{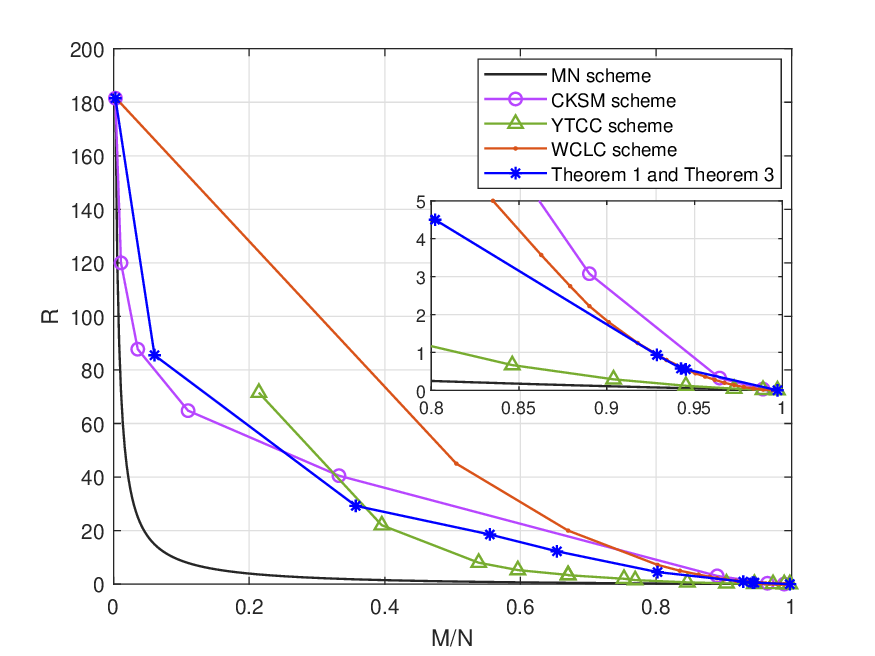}\\
\caption{Transmission load versus memory ratio for $K=364$.}\label{Fig_R_K364}
\end{figure}

\begin{figure}[!htbp]
\centering
\includegraphics[width=4in]{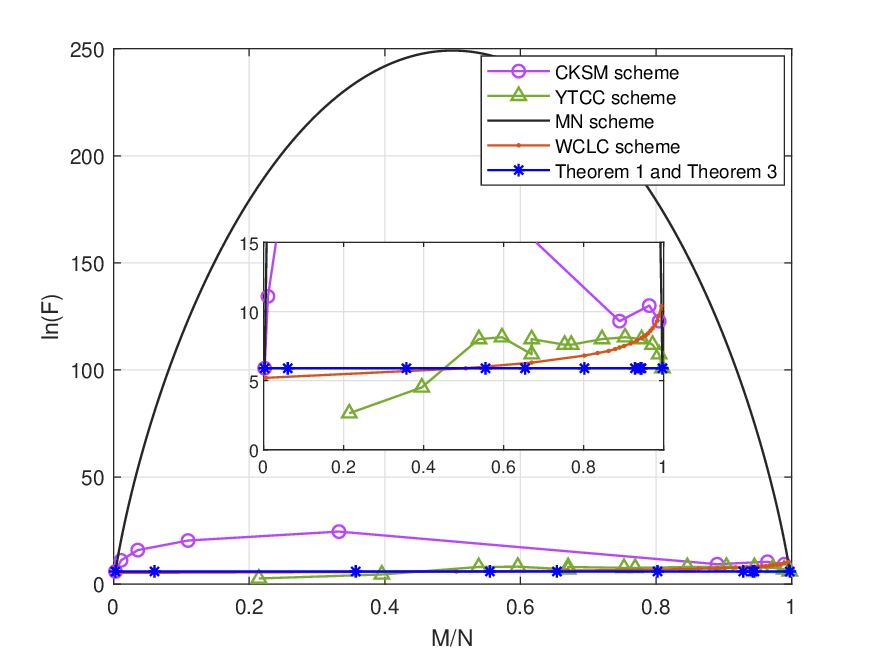}\\
\caption{Natural logarithm of subpacketization versus memory ratio for $K=364$.}\label{Fig_lnF_K364}
\end{figure}

Figs.~\ref{Fig_R_K364} and \ref{Fig_lnF_K364} jointly compare the transmission load and subpacketization of the proposed schemes in Theorems~1 and~3 with those of the MN~\cite{MN}, CKSM~\cite{CKSM}, WCLC~\cite{WCLC}, and YTCC~\cite{YTCC} schemes for \(K=364\). The MN scheme achieves the lowest transmission load but incurs an extremely large subpacketization, with \(\ln(F)\) reaching approximately \(250\) at intermediate memory ratios. In contrast, the proposed schemes maintain a constant subpacketization of \(F=K=364\), i.e., \(\ln(F)\approx5.90\), while achieving competitive transmission loads. Compared with CKSM, the proposed schemes achieve lower transmission loads over a range of medium-to-high memory ratios while substantially reducing subpacketization. Compared with WCLC, they achieve lower transmission loads over a broad range of memory ratios with comparable subpacketization. The proposed schemes also offer a competitive tradeoff between transmission load and subpacketization compared with YTCC. Although YTCC achieves lower transmission loads in some memory regimes, the proposed schemes achieve smaller subpacketization over a considerable range of memory ratios. 

\section{Conclusion}\label{sec:conclusion}
In this paper, we investigated the construction of coded caching schemes with linear subpacketization, particularly \(F=K\). Motivated by the observation that the regularity constraint of conventional PDA-based schemes may limit the achievable transmission performance, we proposed two classes of PDA constructions based on Hamming distance and linear codes. The key idea is to exploit the structures of different Hamming-distance classes to accommodate different coded multicast gains, thereby relaxing the regularity constraint of existing constructions.

Theoretical and numerical comparisons demonstrate the performance advantages of the proposed schemes. For the same number of users and at the same or comparable memory ratios, the proposed schemes can achieve lower transmission loads than several existing linear-subpacketization schemes, including AST, ZCW, WCCLS, ASK, CWWC, and YCWC. Compared with schemes having other subpacketization levels, the proposed schemes can achieve both lower transmission loads and smaller subpacketization than CKSM over a wide range of memory ratios. They can also  achieve lower transmission loads than WCLC while maintaining comparable subpacketization. These results highlight the potential of nonregular PDA constructions for improving the tradeoff between transmission load and subpacketization.

\appendices

\section{Proof of Theorem~\ref{thm:one}}\label{app:one}
\begin{proof}
By Construction~\ref{con:one}, the row and column index sets of $\mathbf{P}$ are both $\mathcal{X} = \prod_{i = 1}^{m}\left\lbrack 0:q_{i} - 1 \right\rbrack$, giving $K = F = |\mathcal{X}| = \prod_{i = 1}^{m}q_{i}$. Next we verify the PDA conditions and derive the other parameters.

\subsubsection{Number of Stars and Condition C1}

Fix a column with index $\mathbf{b} \in \mathcal{X}$. The entry at $(\mathbf{a},\mathbf{b})$ is non-star if and only if \mbox{$d(\mathbf{a},\mathbf{b}) = w$}. Let the set of differing coordinates be $\mathcal{D}=\{i\in[m]:a_i\ne b_i\}$, then $|\mathcal{D}|=w.$ For a fixed $\mathcal{D}$, consider each coordinate of the row index $\mathbf{a}$.
If $i \in \mathcal{D}$, then $a_{i} \neq b_{i}$, giving $q_{i} - 1$ choices for $a_{i}$ among the $q_{i}$ symbols.
If $i \in \mathcal{D}^{c}$, then \mbox{$a_{i} = b_{i}$}, so $a_{i}$ has only one choice.
Thus, $\mathcal{D}$ gives $\prod_{i \in \mathcal{D}}(q_i-1)$ row indices. Distinct sets $\mathcal{D}$ give distinct non-star positions, so the number of non-star entries in the column is \mbox{$\sum_{\mathcal{D} \subseteq \lbrack m\rbrack, |\mathcal{D}| = w}{\prod_{i \in \mathcal{D}}(q_i-1)}.$}
This count is independent of $\mathbf{b}$. Every column therefore contains \mbox{$Z = \prod_{i = 1}^{m}q_{i} - \sum_{\mathcal{D} \subseteq \lbrack m\rbrack, |\mathcal{D}| = w}{\prod_{i \in \mathcal{D}}(q_i-1)}$} stars, proving Condition {\bf C1}.

\subsubsection{Condition C3}

Consider distinct positions $(\mathbf{a},\mathbf{b})$ and $\left( \mathbf{a}',\mathbf{b}' \right)$ satisfying $\mathbf{P}(\mathbf{a},\mathbf{b}) = \mathbf{P}\left( \mathbf{a}',\mathbf{b}' \right) \neq \ast $, which implies both positions have the same set of differing coordinates $\mathcal{D}$, with \mbox{$|\mathcal{D}| = w$}, and \mbox{${\mathbf{a}}_{\mathcal{D}} \cup {\mathbf{b}}_{\mathcal{D}} = \mathbf{a}'_{\mathcal{D}} \cup \mathbf{b}'_{\mathcal{D}},$} \mbox{$\boldsymbol{\xi} + \mathcal{C} = \boldsymbol{\xi}' + \mathcal{C}$} from \eqref{eq:construction-one}.

Let $\lambda = d\left( \boldsymbol{\xi},\boldsymbol{\xi}' \right)$, then the two binary vectors differ in exactly $\lambda$ coordinates. For each coordinate in $\mathcal{D}$, equal binary components correspond to the same ordering of the symbol pair, whereas different components correspond to reversed orderings.
More precisely, write $\mathcal{D}$ as $\mathcal{D} = \{ i_{1},\ldots,i_{w}\}$ in increasing order. For each $j \in \lbrack w\rbrack$,
\begin{equation*}(a'_{i_j},b'_{i_j})=
	\begin{cases}
		(a_{i_j},b_{i_j}),&\xi'_j=\xi_j,\\
		(b_{i_j},a_{i_j}),&\xi'_j\ne\xi_j.
\end{cases}\end{equation*}
Thus, each cross position has exactly $w - \lambda$ differing coordinates in $\mathcal{D}$, i.e., 
\begin{equation}
	\label{eq_ab_D}
	d( {\mathbf{a}}_{\mathcal{D}},\mathbf{b}'_{\mathcal{D}} ) = d( \mathbf{a}'_{\mathcal{D}},{\mathbf{b}}_{\mathcal{D}} ) = w - \lambda.
\end{equation}
We distinguish two cases.

\textbf{Case 1}: \mbox{$\lambda = 0$}. Then $\boldsymbol{\xi} = \boldsymbol{\xi}'$, and equality of the unordered pairs (i.e., ${\mathbf{a}}_{\mathcal{D}} \cup {\mathbf{b}}_{\mathcal{D}} = \mathbf{a}'_{\mathcal{D}} \cup \mathbf{b}'_{\mathcal{D}}$) gives $\mathbf{a}_{\mathcal{D}}=\mathbf{a}'_{\mathcal{D}},\  \mathbf{b}_{\mathcal{D}}=\mathbf{b}'_{\mathcal{D}}.$ On $\mathcal{D}^{c}$, we have ${\mathbf{a}}_{\mathcal{D}^{c}} = {\mathbf{b}}_{\mathcal{D}^{c}}$ and $\mathbf{a}'_{\mathcal{D}^{c}} = \mathbf{b}'_{\mathcal{D}^{c}}$. Since $(\mathbf{a},\mathbf{b})$ and $\left( \mathbf{a}',\mathbf{b}' \right)$ are distinct, ${\mathbf{a}}_{\mathcal{D}^{c}} \neq \mathbf{a}'_{\mathcal{D}^{c}}$ must hold. Therefore, $d\left( {\mathbf{a}}_{\mathcal{D}^{c}},\mathbf{a}'_{\mathcal{D}^{c}} \right) > 0.$ Consequently,
$d(\mathbf{a},\mathbf{b}')=d(\mathbf{a}_{\mathcal{D}},\mathbf{b}'_{\mathcal{D}})+d(\mathbf{a}_{\mathcal{D}^c},\mathbf{b}'_{\mathcal{D}^c})=w+d(\mathbf{a}_{\mathcal{D}^c},\mathbf{a}'_{\mathcal{D}^c})>w$  from \eqref{eq_ab_D}. 
Similarly, $d\left( \mathbf{a}',\mathbf{b} \right) > w$. It implies $\mathbf{P}\left( \mathbf{a},\mathbf{b}' \right) = \mathbf{P}\left( \mathbf{a}',\mathbf{b} \right) = \ast$ from \eqref{eq:construction-one}.

\textbf{Case 2}: $\lambda > 0$. Then $\boldsymbol{\xi} \neq \boldsymbol{\xi}'$. Since the binary vectors belong to the same coset (i.e., $\boldsymbol{\xi} + \mathcal{C} = \boldsymbol{\xi}' + \mathcal{C}$), there exist ${\mathbf{c}_1},{\mathbf{c}}_{2} \in \mathcal{C}$ such that \mbox{$\boldsymbol{\xi} + {\mathbf{c}}_{1} = \boldsymbol{\xi}' + {\mathbf{c}}_{2},$}, i.e., \mbox{$\boldsymbol{\xi} - \boldsymbol{\xi}' = {\mathbf{c}}_{2} - {\mathbf{c}}_{1},$} where ${\mathbf{c}}_{1} \neq {\mathbf{c}}_{2}$. The minimum distance of linear code $\mathcal{C}$ gives
$\lambda = d\left( \boldsymbol{\xi},\boldsymbol{\xi}' \right) = d\left( {\mathbf{c}}_{1},{\mathbf{c}}_{2} \right) \geq m - w + 1.$
Hence,
$d(\mathbf{a},\mathbf{b}')=d(\mathbf{a}_{\mathcal{D}},\mathbf{b}'_{\mathcal{D}})+d(\mathbf{a}_{\mathcal{D}^c},\mathbf{b}'_{\mathcal{D}^c})\le w-\lambda+(m-w)=m-\lambda\le w-1$ from \eqref{eq_ab_D}. 
Similarly, $d\left( \mathbf{a}',\mathbf{b} \right) \leq w - 1$. Thus $\mathbf{P}\left( \mathbf{a},\mathbf{b}' \right) = \mathbf{P}\left( \mathbf{a}',\mathbf{b} \right) = \ast$ from \eqref{eq:construction-one}.

Furthermore, $\mathbf{P}\left( \mathbf{a},\mathbf{b}' \right) = \mathbf{P}\left( \mathbf{a}',\mathbf{b} \right) = \ast$ implies $\mathbf{a} \neq \mathbf{a}'$ and $\mathbf{b} \neq \mathbf{b}'$, proving Condition {\bf C3}.

\subsubsection{Number of Non-Star Symbols and Condition C2}

A non-star symbol in Construction~\ref{con:one} is determined by the label $\left( \mathcal{D},{\mathbf{a}}_{\mathcal{D}} \cup {\mathbf{b}}_{\mathcal{D}},\boldsymbol{\xi} + \mathcal{C} \right)$. Fix $\mathcal{D}$ of size $w$. For each $i \in \mathcal{D}$, choosing two distinct symbols from the $q_{i}$ symbols gives $\binom{q_{i}}{2}$ unordered pairs. The total number of pair collections is $\prod_{i \in \mathcal{D}}\binom{q_{i}}{2}$.
Since $\mathcal{C}$ has length $w$ and dimension $\delta$, it has $2^{w - \delta}$ cosets in $\mathbb{F}_{2}^{w}$. Thus, the number of non-star symbols for $\mathcal{D}$ is $2^{w - \delta}\prod_{i \in \mathcal{D}}\binom{q_{i}}{2}$. Summing over all sets $\mathcal{D}$ of size $w$ gives $$\textstyle S = 2^{w - \delta}\sum_{\mathcal{D} \subseteq \lbrack m\rbrack, |\mathcal{D}| = w}{\prod_{i \in \mathcal{D}}\binom{q_{i}}{2}}.$$

Hence, the array \(\mathbf P\) generated by Construction~\ref{con:one} is a PDA with the parameters specified in Theorem~\ref{thm:one}. The corresponding coded caching parameters are then obtained using \(M/N=Z/F\) and \(R=S/F\) established in Lemma~\ref{lem:pda}.
\end{proof}

\section{Proof of Lemma~\ref{lemm1}}\label{app_lemm1}
\begin{proof}
The array generated by Construction \ref{constr2} has $F=q^m$ rows and $K=q^m$ columns, with each column containing $Z=q^m-\binom{m}{w}(q-1)^w$ stars. 
For any two distinct positions
$(\mathbf{a},\mathbf{b})$ and $(\mathbf{a}',\mathbf{b}')$, assume that
$
\mathbf{P}(\mathbf{a},\mathbf{b})
=
\mathbf{P}(\mathbf{a}',\mathbf{b}')
\neq *$, we will prove that $\mathbf{P}(\mathbf{a},\mathbf{b}')
=
\mathbf{P}(\mathbf{a}',\mathbf{b})=*$.
From \eqref{eq:construction2} we have $	\left(
\mathcal{D},\,
\mathbf{a}_{\mathcal{D}}\cup\mathbf{b}_{\mathcal{D}},\,
\mathbf{a}_{\mathcal{D}^{c}}+\mathcal{C}'
\right)
=
\left(
\mathcal{D},\,
\mathbf{a}'_{\mathcal{D}}\cup\mathbf{b}'_{\mathcal{D}},\,
\mathbf{a}'_{\mathcal{D}^{c}}+\mathcal{C}'
\right)$, where $\mathcal{D}=\{\,i\in[m]\mid a_i\neq b_i\,\}
=
\{\,i\in[m]\mid a_i'\neq b_i'\,\}$. 	
Hence,
\begin{align}		
	\mathbf{a}_{\mathcal{D}}\cup\mathbf{b}_{\mathcal{D}}
	&=
	\mathbf{a}'_{\mathcal{D}}\cup\mathbf{b}'_{\mathcal{D}}, \label{eq_union2}\\
	\mathbf{a}_{\mathcal{D}^{c}}+\mathcal{C}'
	&=
	\mathbf{a}'_{\mathcal{D}^{c}}+\mathcal{C}'.\label{eq_xic2}
\end{align}
There are two cases.

\emph{Case 1}: If \(\mathbf{a}_{\mathcal{D}^{c}}=\mathbf{a}'_{\mathcal{D}^{c}}\), we have $d(\mathbf{a}_{\mathcal{D}},\mathbf{b}'_{\mathcal{D}})<w$. Otherwise, we have $\mathbf{a}_{\mathcal{D}}=\mathbf{a}'_{\mathcal{D}}$ and $\mathbf{b}_{\mathcal{D}}=\mathbf{b}'_{\mathcal{D}}$ from \eqref{eq_union2}. Then $\mathbf{a}=\mathbf{a}'$ and $\mathbf{b}=\mathbf{b}'$,
which contradicts the assumption that $(\mathbf{a},\mathbf{b})$ and $(\mathbf{a}',\mathbf{b}')$ are distinct positions. Therefore, $d(\mathbf{a},\mathbf{b}')=
d(\mathbf{a}_{\mathcal{D}},\mathbf{b}'_{\mathcal{D}})
+
d(\mathbf{a}_{\mathcal{D}^{c}},\mathbf{b}'_{\mathcal{D}^{c}})=d(\mathbf{a}_{\mathcal{D}},\mathbf{b}'_{\mathcal{D}})
+
d(\mathbf{a}_{\mathcal{D}^{c}},\mathbf{a}'_{\mathcal{D}^{c}})=d(\mathbf{a}_{\mathcal{D}},\mathbf{b}'_{\mathcal{D}})<w$, which implies
$\mathbf{P}(\mathbf{a},\mathbf{b}')=*$ from \eqref{eq:construction2}.

\emph{Case 2}: If \(\mathbf{a}_{\mathcal{D}^{c}}\neq\mathbf{a}'_{\mathcal{D}^{c}}\), $\mathcal{C}'$ is not the zero code from \eqref{eq_xic2}, which implies $2w+1\leq m$. Moreover, 
there exist
$\mathbf{c}_1,\mathbf{c}_2\in \mathcal{C}'$
such that
$\mathbf{a}_{\mathcal{D}^{c}}+\mathbf{c}_1
=
\mathbf{a}'_{\mathcal{D}^{c}}+\mathbf{c}_2$, leading to $\mathbf{a}_{\mathcal{D}^{c}}-\mathbf{a}'_{\mathcal{D}^{c}}
=
\mathbf{c}_2-\mathbf{c}_1$.
Hence,
$d(\mathbf{a}_{\mathcal{D}^{c}},\mathbf{a}'_{\mathcal{D}^{c}})
=
d(\mathbf{c}_1,\mathbf{c}_2)\geq w+1$ from the Hamming distance constraint of  \(\mathcal{C}'\). Therefore, $d(\mathbf{a},\mathbf{b}')=
d(\mathbf{a}_{\mathcal{D}},\mathbf{b}'_{\mathcal{D}})
+
d(\mathbf{a}_{\mathcal{D}^{c}},\mathbf{b}'_{\mathcal{D}^{c}})=d(\mathbf{a}_{\mathcal{D}},\mathbf{b}'_{\mathcal{D}})
+
d(\mathbf{a}_{\mathcal{D}^{c}},\mathbf{a}'_{\mathcal{D}^{c}})\geq w+1$, which implies
$\mathbf{P}(\mathbf{a},\mathbf{b}')=*$ from \eqref{eq:construction2}.

Similarly, one can also prove $\mathbf{P}(\mathbf{a}',\mathbf{b})=*$.

Next we will prove that each non-star symbol appears exactly $2^{w}\nu$ times.
For any non-star entry 
$\mathbf{P}(\mathbf{a},\mathbf{b})
=
\left(
\mathcal{D},\,
\mathbf{a}_{\mathcal{D}}\cup\mathbf{b}_{\mathcal{D}},\,
\mathbf{a}_{\mathcal{D}^{c}}+\mathcal{C}'
\right)$,
and for any
$\mathbf{c}\in\mathbf{a}_{\mathcal{D}^{c}}+\mathcal{C}'$, let $\mathbf{a}'_{\mathcal{D}^{c}}=\mathbf{b}'_{\mathcal{D}^{c}}=\mathbf{c}$. Let $\mathbf{a}'_{\mathcal{D}}, \mathbf{b}'_{\mathcal{D}}\in [0:q-1]^w$ such that $\mathbf{a}'_{\mathcal{D}}\cup\mathbf{b}'_{\mathcal{D}}=\mathbf{a}_{\mathcal{D}}\cup\mathbf{b}_{\mathcal{D}}$. Then we have $\mathbf{P}(\mathbf{a}',\mathbf{b}')
=\mathbf{P}(\mathbf{a},\mathbf{b})$ from \eqref{eq:construction2}. Since $|\mathcal{C}'|=\nu$ and there are $2^w$ choices for $\mathbf{a}'_{\mathcal{D}}, \mathbf{b}'_{\mathcal{D}}$, each non-star symbol appears $2^w\nu$ times in $\mathbf{P}$. Thus, there are $S=\frac{q^m\binom{m}{w}(q-1)^w}{2^{w}\nu}$ different non-star symbols in $\mathbf{P}$.
Hence, the array generated by Construction \ref{constr2} is a $2^w\nu$-$\big(q^m,\,q^m,\,q^m-\binom{m}{w}(q-1)^w,\,\frac{q^m\binom{m}{w}(q-1)^w}{2^w\nu}
\big)$ PDA.
\end{proof}

\section{Proof of Theorem~\ref{thm:two}}\label{app:two}
\begin{proof}
By Construction~\ref{con:two}, the row and column index sets of $\mathbf{P}$ are both $\mathcal{X} = \prod_{i = 1}^{m}\left\lbrack 0:q_{i} - 1 \right\rbrack$, so \mbox{$K = F = \prod_{i = 1}^{m}q_{i}$}. Next we verify the PDA conditions and derive the parameters.

\subsubsection{Number of Stars and Condition C1}

Fix a column with index $\mathbf{b} \in \mathcal{X}$. The entry at $(\mathbf{a},\mathbf{b})$ is non-star if and only if $d(\mathbf{a},\mathbf{b}) \geq w$.
For each $r \in \lbrack w:m\rbrack$, fix a set of differing coordinates $\mathcal{D}$ of size $r$. If $i \in \mathcal{D}$, $a_{i}$ has $q_{i} - 1$ choices; if $i \in \mathcal{D}^{c}$, then \mbox{$a_{i} = b_{i}$}. The number of such row indices is $\prod_{i \in \mathcal{D}}(q_i-1)$.
Summing over all sets $\mathcal{D}$ of size $r$ and then over $r \in \lbrack w:m\rbrack$ gives the number of non-star entries per column: \mbox{$\sum_{r = w}^{m}{\sum_{\mathcal{D} \subseteq \lbrack m\rbrack, |\mathcal{D}| = r}{\prod_{i \in \mathcal{D}}(q_i-1)}}.$}
This count is independent of the column index. Therefore, there are $$\textstyle Z = \prod_{i = 1}^{m}q_{i} - \sum_{r = w}^{m}{\sum_{\mathcal{D} \subseteq \lbrack m\rbrack, |\mathcal{D}| = r}{\prod_{i \in \mathcal{D}}(q_i-1)}}$$ stars in each column, so {\bf C1} holds.

\subsubsection{Condition C3}

Consider distinct positions $(\mathbf{a},\mathbf{b})$ and $\left( \mathbf{a}',\mathbf{b}' \right)$ satisfying $\mathbf{P}(\mathbf{a},\mathbf{b}) = \mathbf{P}\left( \mathbf{a}',\mathbf{b}' \right) \neq \ast .$ It implies they have the same set of differing coordinates $\mathcal{D}$ from \eqref{eq:construction-two}. Let \mbox{$r = |\mathcal{D}|$}, then $r \geq w$, and \mbox{${\mathbf{a}}_{\mathcal{D}} \cup {\mathbf{b}}_{\mathcal{D}} = \mathbf{a}'_{\mathcal{D}} \cup \mathbf{b}'_{\mathcal{D}},$} \mbox{${\mathbf{a}}_{\mathcal{D}^{c}} = \mathbf{a}'_{\mathcal{D}^{c}},$} \mbox{$\boldsymbol{\xi} + \mathcal{C}_{r} = \boldsymbol{\xi}' + \mathcal{C}_{r}.$}
Since $\mathcal{D}^{c}$ consists of the coordinates where the row and column indices agree, then ${\mathbf{a}}_{\mathcal{D}^{c}} = {\mathbf{b}}_{\mathcal{D}^{c}}=\mathbf{a}'_{\mathcal{D}^{c}} = \mathbf{b}'_{\mathcal{D}^{c}}$. So
\begin{equation}
	\label{eq_dab'Dc}
	d(\mathbf{a}_{\mathcal{D}^{c}},\mathbf{b}'_{\mathcal{D}^{c}})=d(\mathbf{a}'_{\mathcal{D}^{c}},\mathbf{b}_{\mathcal{D}^{c}})=0. 
\end{equation}

Let $\lambda = d\left( \boldsymbol{\xi},\boldsymbol{\xi}' \right)$. If \mbox{$\lambda = 0$}, then $\boldsymbol{\xi} = \boldsymbol{\xi}'$, and equality of the unordered pairs (i.e., ${\mathbf{a}}_{\mathcal{D}} \cup {\mathbf{b}}_{\mathcal{D}} = \mathbf{a}'_{\mathcal{D}} \cup \mathbf{b}'_{\mathcal{D}}$) gives ${\mathbf{a}}_{\mathcal{D}} = \mathbf{a}'_{\mathcal{D}}$ and ${\mathbf{b}}_{\mathcal{D}} = \mathbf{b}'_{\mathcal{D}}$. The common values on $\mathcal{D}^{c}$ (i.e., ${\mathbf{a}}_{\mathcal{D}^{c}} = \mathbf{a}'_{\mathcal{D}^{c}}$) then imply \mbox{$\mathbf{a} = \mathbf{a}'$} and \mbox{$\mathbf{b} = \mathbf{b}'$}, contradicting the distinctness of the positions. Hence, $\lambda > 0$.

Since the cosets agree (i.e., $\boldsymbol{\xi} + \mathcal{C}_{r} = \boldsymbol{\xi}' + \mathcal{C}_{r}$), there exist distinct ${\mathbf{c}}_{1},{\mathbf{c}}_{2} \in \mathcal{C}_{r}$ such that \mbox{$\boldsymbol{\xi} + {\mathbf{c}}_{1} = \boldsymbol{\xi}' + {\mathbf{c}}_{2},$} thus \mbox{$\boldsymbol{\xi} - \boldsymbol{\xi}' = {\mathbf{c}}_{2} - {\mathbf{c}}_{1}.$} 
The distance condition of $\mathcal{C}_{r}$ gives
\begin{equation}
	\label{eq_lambda}
	\lambda = d\left( \boldsymbol{\xi},\boldsymbol{\xi}' \right) = d\left( {\mathbf{c}}_{1},{\mathbf{c}}_{2} \right) \geq r - w + 1.
\end{equation}
Write $\mathcal{D}$ as $\mathcal{D} = \{ i_{1},\ldots,i_{r}\}$ in increasing order. Since ${\mathbf{a}}_{\mathcal{D}} \cup {\mathbf{b}}_{\mathcal{D}} = \mathbf{a}'_{\mathcal{D}} \cup \mathbf{b}'_{\mathcal{D}}$, for each $j \in \lbrack r\rbrack$, from \eqref{eq_xi2} we have
\begin{equation*}(a'_{i_j},b'_{i_j})=
	\begin{cases}
		(a_{i_j},b_{i_j}),&\xi'_j=\xi_j,\\
		(b_{i_j},a_{i_j}),&\xi'_j\ne\xi_j.
\end{cases}\end{equation*}
Hence, \mbox{$d( {\mathbf{a}}_{\mathcal{D}},\mathbf{b}'_{\mathcal{D}} ) = d( \mathbf{a}'_{\mathcal{D}},{\mathbf{b}}_{\mathcal{D}} ) = r - \lambda.$}
Then from \eqref{eq_dab'Dc} and \eqref{eq_lambda} we have
$d(\mathbf{a},\mathbf{b}')=d(\mathbf{a}_{\mathcal{D}},\mathbf{b}'_{\mathcal{D}})+d(\mathbf{a}_{\mathcal{D}^c},\mathbf{b}'_{\mathcal{D}^c})=r-\lambda\le r-(r-w+1)=w-1$, which implies $\mathbf{P}\left( \mathbf{a},\mathbf{b}' \right) =\ast$ from \eqref{eq:construction-two}. 
Similarly, we can prove $\mathbf{P}\left( \mathbf{a}',\mathbf{b} \right) = \ast$.

Moreover, $\lambda > 0$ guarantees at least one coordinate with different binary components. At that coordinate, both the row and column indices differ. Thus, $\mathbf{a} \neq \mathbf{a}'$ and $\mathbf{b} \neq \mathbf{b}'$, proving {\bf C3}.

\subsubsection{Number of Non-Star Symbols and Condition C2}

A non-star symbol in Construction~\ref{con:two} is determined by the label $\left( \mathcal{D},{\mathbf{a}}_{\mathcal{D}} \cup {\mathbf{b}}_{\mathcal{D}},{\mathbf{a}}_{\mathcal{D}^{c}},\boldsymbol{\xi} + \mathcal{C}_{r} \right)$, where \mbox{$r = |\mathcal{D}|$}. Fix a distance $r \in \lbrack w:m\rbrack$ and a set $\mathcal{D}$ of size $r$. The other three components have the following numbers of choices: 
\begin{itemize}
	\item There are $\prod_{i \in \mathcal{D}}\binom{q_{i}}{2}$ collections of unordered symbol pairs ${\mathbf{a}}_{\mathcal{D}} \cup {\mathbf{b}}_{\mathcal{D}}$.
	\item There are $\prod_{i \in \mathcal{D}^{c}}q_{i}$ choices of common values on $\mathcal{D}^{c}$.
	\item The code $\mathcal{C}_{r}$ has length $r$ and dimension $\delta_r$, giving $2^{r - \delta_r}$ cosets.
\end{itemize}
Thus, the number of non-star symbols for $\mathcal{D}$ is
$$\textstyle 2^{r - \delta_r}\prod_{i \in \mathcal{D}}\binom{q_{i}}{2}\prod_{i \in \mathcal{D}^{c}}q_{i}=\frac{\prod_{i=1}^m q_i}{2^{\delta_r}}\prod_{i\in \mathcal{D}}(q_i-1).$$
Summing over all $\mathcal{D}$ and $r$ gives
$$\textstyle S = \prod_{i = 1}^{m}q_{i}\sum_{r = w}^{m}\frac{1}{2^{\delta_r}}\sum_{\mathcal{D} \subseteq \lbrack m\rbrack, |\mathcal{D}| = r}{\prod_{i \in \mathcal{D}}(q_i-1)}.$$

Therefore, the array generated by Construction~\ref{con:two} is a PDA with the parameters specified in Theorem~\ref{thm:two}. The corresponding coded caching parameters are obtained using \(M/N=Z/F\) and \(R=S/F\) given in Lemma~\ref{lem:pda}.
\end{proof}

\bibliographystyle{IEEEtran}
\bibliography{reference}

\end{document}